\documentclass[journal,10pt,twocolumn]{IEEEtran}
\usepackage{epsfig}
\usepackage{subfigure}
\usepackage{graphicx}
\usepackage{times}
\usepackage{cite}
\usepackage{enumerate}
\usepackage{amssymb}
\usepackage{algorithm}
\usepackage{algorithmic}
\usepackage{multirow}
\usepackage{amsmath}
\usepackage{braket}

\newtheorem{example}{Example}[section]
\newtheorem{remark}{Remark}[section]

\begin{document}

\title{Neighbor Discovery in a Wireless Sensor Network: Multipacket Reception Capability and Physical-Layer Signal Processing}


\author{
\authorblockN{Jeongho Jeon and Anthony Ephremides}
%
\thanks{The material in this paper was presented in part at the 48th Annual Allerton Conference on Communication, Control, and Computing (Monticello, IL), Sept. 2010.}
\thanks{The authors are with the
Department of Electrical and Computer Engineering and the Institute
for Systems Research, University of Maryland, College Park, MD 20742
USA (e-mail: jeongho@umd.edu; etony@umd.edu).}}
\maketitle

\begin{abstract}
In randomly deployed networks, such as sensor networks, an important
problem for each node is to discover its \textit{neighbor} nodes so
that the connectivity amongst nodes can be established. In this
paper, we consider this problem by incorporating the physical layer
parameters in contrast to the most of the previous work which
assumed a collision channel. Specifically, the pilot signals that
nodes transmit are successfully decoded if the strength of the
received signal relative to the interference is sufficiently high.
Thus, each node must extract signal parameter information from the
superposition of an unknown number of received signals. This problem
falls naturally in the purview of random set theory (RST) which
generalizes standard probability theory by assigning \textit{sets},
rather than values, to random outcomes. The contributions in the
paper are twofold: first, we introduce the realistic effect of
physical layer considerations in the evaluation of the performance
of \textit{logical} discovery algorithms; such an introduction is
necessary for the accurate assessment of how an algorithm performs.
Secondly, given the \textit{double} uncertainty of the environment
(that is, the lack of knowledge of the number of neighbors along
with the lack of knowledge of the individual signal parameters), we
adopt the viewpoint of RST and demonstrate its advantage relative to
classical matched filter detection method.
\end{abstract}


\section{Introduction}

\PARstart{W}{ireless} sensor networks are gaining great attention
due to its versatility in civilian/military applications such as
environmental monitoring and target detection/tracking in a
cost-effective manner. In such applications, a large number of
sensors are randomly deployed over the region of interest and,
presumably, neighbor discovery is the first and foremost process to
run after the deployment to form a \textit{network} whose
connectivity greatly affects the performance of subsequent network
operations over the entire life span. The challenge is compounded by
the fact that neighbor discovery has to be done without any a priori
knowledge on the random deployment or any communication
infrastructures.
%
%

Neighbor discovery in wireless networks is defined to be the process
to identify a set of nodes with which a node can communicate, and it
has been addressed by several authors \cite{borbash:asynchronous,
vasudevan:coupon, vasudevan:neighbor, mcglynn:birthday}. In
\cite{borbash:asynchronous}, a simple ALOHA-like neighbor discovery
algorithm was proposed in which each node randomly transmit/listen
in each time slot and analyzed for both synchronous and asynchronous
timing cases. This type of discovery algorithm based on the random
access protocol is well suited for randomly distributed wireless
networks. In \cite{vasudevan:coupon}, similar neighbor discovery
algorithm was considered and the expected time to find neighbors was
obtained. In \cite{vasudevan:neighbor}, a \textit{gossip-based
algorithm} was proposed in which each node transmits a table of
\textit{gossip} data (which is the list of neighbors that it has
discovered so far and their locations) in a random direction using
directional antennas. In \cite{mcglynn:birthday}, a family of
probabilistic protocols, called birthday protocols, have been
proposed to initiate the randomly deployed wireless networks. From a
physical-layer point of view, however, the previous works are
extremely limited due to the use of collision channel model. Under
this model, if more than one nodes transmit at the same time, none
of them are successful. However, it is too pessimistic in the sense
that a transmission may succeed even in the presence of interference
which is called \textit{capture effect} \cite{zorzi:capture,
nguyen:capture, namislo:analysis, hajek:capture, ghez:stability}.
We, thus, claim that the performance of neighbor discovery algorithm
has been quite underestimated so far due to the use of unrealistic
channel model, and correct reassessment of the discovery algorithm
is required.


In this paper, we consider a shared channel and nodes with
multipacket reception capability in which a transmission is
successful if the received signal-to-interference-plus-noise-ratio
(SINR) exceeds a certain threshold. Specifically, under the
discovery algorithm proposed in \cite{borbash:asynchronous}, we
obtain the expression for the expected number of successful
receptions \textit{per slot} at a given SINR threshold and find the
optimal transmission probability which maximizes the expected number
of successful receptions. We note, however, that for a given
modulation scheme and target bit error rate (BER), the data rate is
an increasing function of the SINR threshold. Therefore, for a fair
comparison, we normalize the slot duration as the unit time and
obtain the expected number of successful receptions \textit{per
second}. After that, the performance of the discovery algorithm over
multiple slots is studied which is useful, for example, in deciding
the duration of the discovery process.
%

In dealing with multipacket reception capability, the physical layer
signal processing issues cannot be overlooked. This is because the
received signal at each time slot is the superposition of the
signals transmitted from random set of nodes and noise. Therefore,
determining the existence of a signal from a particular node itself
is not an easy task. For this problem, we first present the
classical matched filter method which fundamentally treats the
interference as noise. As an alternative, a more accurate method can
be envisioned in an additional cost of complexity. Since the number
of transmitters and their entities are all unknown, we adopt the
viewpoint of random set theory (RST: see Appendix \ref{appendix:rst}
and references therein) and propose RST-based method for detecting
the transmitting nodes in each time slot \cite{goodman:mathematics,
ali:random, vihola:random, mahler:engineering, mahler:random,
mahler:statistical, biglieri:multiuser}. Besides, it is also
possible with RST to estimate additional parameters of transmitted
signals such as signal amplitudes and phase.


This paper is organized as follows. The next section describes the
basic assumptions, establishes the notation, and presents the signal
model. In Section \ref{sec:algorithm}, details on the neighbor
discovery algorithm is presented and the early termination of the
discovery process is discussed. In Section \ref{sec:analysis_mpr},
we analyze the chosen discovery algorithm under the SINR-based
model. In Section \ref{sec:matched_filter}, physical layer signal
processing solutions are delivered. Specifically, we obtain matched
filter detector for neighbor discovery and, as an alternative,
RST-based maximum a posteriori (MAP) estimator is proposed. In
Section \ref{sec:numerical_results}, numerical results are
presented.
%
%
%
Finally, we draw conclusions in Section \ref{secCon}.

\section{System Model}\label{sec:system}

We consider a time-slotted wireless sensor network which is deployed
over a region of interest such as large tactical area for target
detection or vast rural area for environmental observation. In such
scenarios, a large number of sensor nodes are released from an
airplane. For a large number of nodes over a large area, the
locations of nodes are modeled by a homogeneous, two-dimensional
Poisson point process with intensity $\lambda$ which is the average
number of nodes per unit area. Therefore, the number of nodes in a
unit area follows Poisson distribution with parameter $\lambda$. As
is well known, a spatial Poisson process on the plane, conditioned
on a given number of nodes within a given area, yields the uniform
distribution of these nodes in that area. As a result, for example,
if a node's discovery region is modeled as a circle of radius $R_0$,
the cumulative probability distribution on the distance from the
node to the other nodes in the region is given by
\begin{equation}\label{eqn:distribution-fn}
\begin{array}{lll}
    F_r(x) \;=\; \left\{ \begin{array}{cl}
                \displaystyle 0,\;& \ \mathrm{if} \ \ x<0\\
                \displaystyle \left(\frac{x}{R_0}\right)^2,\;& \ \mathrm{if} \ \ 0 \leq x \leq R_0\\
                \displaystyle 1,\;& \ \mathrm{if} \ \ x>R_0
                \end{array}\right.

\end{array}
\end{equation}

%
%

The signal received by a node at time slot $t$ (if the node is
listening) is given by
\begin{equation}\label{eqn:received-signal}
\mathbf{y}_t = \sum_{k \in {\mathbf{I}_t}} {g_t^k \mathbf{s}_t^k} +
\mathbf{n}_t
\end{equation}
where $\mathbf{I}_t$ is the set of transmitting neighbors, $g_t^k$
is the complex amplitude of the signal received from the $k$-th
node, $\mathbf{s}_t^k$ is the signal transmitted from the $k$-th
node which is the message multiplied by the signature sequence
bitwise, and $\mathbf{n}_t$ is a random noise. We assume that the
signatures of all nodes are known to each other by assuming that
they share an identical key generator which can be implemented using
a linear feedback shift register. If the length of signature is $L$
and each node transmits a 1-bit known message, then $\mathbf{y}_t =
\{y_{t,1},...,y_{t,L}\}^T$, $\mathbf{s}^k_t =
\{s^k_{t,1},...,s^k_{t,L}\}^T$ which is equal to the signature
sequence, and $\mathbf{n}_t = \{n_{t,1},...,n_{t,L}\}^T$, where the
symbol $T$ denotes vector transpose.
%
%
The noise samples in $\mathbf{n}_t$ are assumed to be independent
and identically distributed (i.i.d.) with the normal distribution
$\mathcal{N}(0,N)$ where $N$ is the noise power which is the product
of noise spectral density $N_0$ and bandwidth $B$. The complex
amplitude $g_t^k$ is of the form $g_t^k=\sqrt{G (1+r_k)^{-\eta}}
\psi_t^k$, where $G$ is the transmission power, $r_k$ is the
distance from the $k$-th node to the receiver, $\eta$ is the path
loss exponent, and $\psi_t^k$ is the channel fading coefficient
which is modeled by i.i.d. standard circular symmetric Gaussian
random process\cite{tse:fundamentals}.

Denote by $\mathcal{J}$ and $J$ the set of neighbors and the number
of neighbors of a node, respectively, and we suppressed the
particular node index for notational brevity. The SINR of node $k$
at time slot $t$ (if the $k$-th node is transmitting) is given by
\begin{equation*}\label{eqn:sinr}
\mathrm{SINR}_t^k = \frac{P_{rx,k}}{\displaystyle \sum_{i \in
{\mathbf{I}_t}, i\neq k} P_{rx,i} + N}
\end{equation*}
where $P_{rx,k}$ is the received signal power from the $k$-th
transmitter. Note that the distances $r_k$ $(k=1,\dots,J)$ are
i.i.d. under the Poisson point process modeling. Also the channel
fading coefficients $\psi^k_t$ are i.i.d., and by further assuming
that $r_k$ and $\psi^k_t$ are mutually independent, the received
signal powers $P_{rx,k}$ $(k=1,\dots,J)$ are i.i.d. as well, and the
common cumulative distribution function of the received signal power
is obtained by \cite{nguyen:capture}
\begin{equation}\label{eqn:cdf_rx_power_1}
\displaystyle F_{P}(x) = \displaystyle 1- \int_{0}^{\infty} F_r
\left( \left(\frac{ \omega G }{x} \right)^{\frac{1}{\eta}} -1
\right) f_{|\psi|^2}(\omega) d\omega
\end{equation}
where $f_{|\psi|^2}(\cdot)$ is the probability density of the
squared magnitude of the stationary fading process which is an
exponential with unit mean. The transmission of the $k$-th node is
said to be successful if
\begin{equation}\label{eqn:sinr-criterion}
\mathrm{SINR}_t^{k} \geq \tau
\end{equation}
where the threshold $\tau$ depends on parameters such as data rate
and target BER \cite{goldsmith:wireless}. Note that, if $\tau<1$, it
is possible for two or more signals to simultaneously satisfy
(\ref{eqn:sinr-criterion}), and if $\tau \geq 1$, at most one signal
having the highest SINR may satisfy (\ref{eqn:sinr-criterion}). If
$\tau$ goes to $\infty$ and we neglect the noise effect, the
criterion in (\ref{eqn:sinr-criterion}) is equivalent to the
collision channel model, and if $\tau$ goes to 0, all transmissions
will be successful, but at the same time, the data rate goes to 0.

\section{The Neighbor Discovery Algorithm}\label{sec:algorithm}

\subsection{Description of the Neighbor Discovery Algorithm}\label{sec:algorithm_description}

An ALOHA-like neighbor discovery algorithm proposed in
\cite{borbash:asynchronous} is reconsidered in which each node
transmits with probability $p_{\mathrm{T}}$ or listens with
probability $1-p_{\mathrm{T}}$. The transmission probability
$p_{\mathrm{T}}$ and the transmission power $G$ are identical for
all nodes; all these simplifying assumptions were made to reduce
non-essential complexities. Notice that a plain method such as the
periodic beaconing can be thought but such a static approach would
certainly fail in a randomly deployed network which has irregular
node density. Under the chosen discovery algorithm, in order for a
node to be discovered by some other node, the former should transmit
and the other should listen at the same time. On top of that, the
transmission must be successful which requires a certain criterion;
in the collision channel, the criterion is that there is only one
transmission at that time slot. However, we adopt a more realistic
SINR criterion for success and optimize the transmission probability
which has been quite underestimated so far due to the use of
collision channel model.

\subsection{Considerations on Early
Termination}\label{sec:protocol:et}

Obviously, the marginal revenue of running the discovery algorithm
diminishes as time goes and, because sensors are usually assumed to
be battery-powered, it is needed to consider an early termination of
the discovery process to prolong the lifetime of the network.
Several metrics can be envisioned as a criterion for the early
termination. A simple criterion is to terminate when a node has
discovered a predetermined number of neighbors. In the case of nodes
located at the sparsely populated areas, however, this may not be
satisfied until the end of the discovery process (even if all the
neighbors are discovered early). Alternatively, one may want to
allow an early termination if a node has discovered a predetermined
fraction of neighbors. However, because each node does not know
\textit{a priori} how many neighbors it has, this criterion is
untestable. As will be seen in the next section, as time goes, the
set of discovered neighbors in each time slot will overlap with the
previously discovered ones. Therefore, if a node do not find any new
neighbors for a sufficiently large number of slots, then it can be
regarded that all the neighbors have been discovered. Hence, an
early termination can be declared if a node does not receive any new
messages during a predetermined number of time slots.

\section{Analysis of the Discovery Algorithm with the Multipacket Reception Capability}\label{sec:analysis_mpr}

\subsection{Optimal Transmission Probability}\label{sec:example}

We start by deriving the expected number of successful receptions in
each time slot as a function of SINR threshold $\tau$. After that
the transmission probability $p_{\mathrm{T}}$ is set to maximize it.
%
%
We first denote by $\mathbf{I}_t^s (\subseteq \mathbf{I}_t)$ the set
of transmitting neighbors satisfying the SINR criterion in
\eqref{eqn:sinr-criterion}. Then, the expected number of successful
receptions by one node is expressed as\footnote{$|\cdot|$ is the
cardinality of a set.}
\begin{align}\label{eqn:capture_num_success}
E[|\mathbf{I}_t^s|] & = \displaystyle \mathrm{Pr}\left\{ \mathrm{a\
\! node\ \! is\ \! listening} \right\}
\displaystyle \sum_{n=1}^{J} \mathrm{Pr}\left\{|\mathbf{I}_t|=n \right\} S_n \nonumber \\
& = \displaystyle \sum_{n=1}^{J} { J \choose n} {p_{\mathrm{T}}}^n
(1-p_{\mathrm{T}})^{J-n+1} S_n
\end{align}
where $S_n$ is the expected number of successful receptions given
$n(\geq 1)$ simultaneous transmissions and is obtained by
\begin{equation}\label{eqn:capture_num_success_given}
    S_n = n \mathrm{Pr} \Set{\mathrm{SINR}_t^1>\tau |\ \!\!
|\mathbf{I}_t|=n, 1 \in \mathbf{I}_t}
\end{equation}
where $\mathrm{SINR}_t^1$ is the SINR of the first transmitter
\cite{nguyen:capture}. Note that
\eqref{eqn:capture_num_success_given} follows from the assumption
that the received signal powers $P_{rx,k}$ $(k=1,\dots,J)$ are
i.i.d. and, thus, the first transmitter needs not be the closest one
to the receiver. It is computed as
\begin{multline}\label{eqn:capture-cal}
\displaystyle \mathrm{Pr} \Set{\mathrm{SINR}_t^1 >\tau |\ \!\! |
\mathbf{I}_t|=n,1 \in \mathbf{I}_t } \\
= \displaystyle 1 - \int_0^{\infty} \cdots  \int_0^{\infty} F_P
\left(\tau \sum_{i=2}^n x_i \right) dF_P(x_2) \cdots dF_P(x_n)
\end{multline}
where the noise effect was neglected for simplicity. Consequently,
in principle, we can find the optimal transmission probability that
maximizes the expected number of successful receptions.

\begin{example}\label{example:opt_tx_prob}

As an illustration of the use of above equations, consider a simple
example in which there are three nodes and they are within the radio
range of each other. Since all three nodes have two neighbors, the
expected number of successful receptions by one node is given by
\begin{multline*}
E[|\mathbf{I}_t^s|] = \displaystyle 2 p_{\mathrm{T}}
(1-p_{\mathrm{T}})^{2} \mathrm{Pr} \Set{\mathrm{SINR}_t^1>\tau |\ \!\! |\mathbf{I}_t| = 1, 1 \in \mathbf{I}_t } \\
+ \displaystyle 2 {p_{\mathrm{T}}}^2 (1-p_{\mathrm{T}})
\mathrm{Pr}\Set{\mathrm{SINR}_t^1>\tau |\ \!\!|\mathbf{I}_t| = 2, 1
\in \mathbf{I}_t }
\end{multline*}
Let us further consider a simplified path loss model in which the
received signal power is given by $P_{rx,k} = r_k^{-\eta}$. Then, it
can be easily shown that for $\tau < 1$,
\begin{equation}\label{eqn:num-success-reception-n2-tau1}
E[|\mathbf{I}_t^s|] = \displaystyle \tau^{\frac{2}{\eta}}
{p_{\mathrm{T}}}^3 - \left( 2 + \tau^{\frac{2}{\eta}} \right)
{p_{\mathrm{T}}}^2 + 2p_{\mathrm{T}}
\end{equation}
and for $\tau \geq 1$,
ff\begin{equation}\label{eqn:num-success-reception-n2-tau2}
E[|\mathbf{I}_t^s|] = \displaystyle \left( 2 -
\tau^{-\frac{2}{\eta}} \right) {p_{\mathrm{T}}}^3 + \left(
\tau^{-\frac{2}{\eta}} -4 \right) {p_{\mathrm{T}}}^2 +
2p_{\mathrm{T}}
\end{equation}
Note that (\ref{eqn:num-success-reception-n2-tau1}) and
(\ref{eqn:num-success-reception-n2-tau2}) are (strictly) concave
over the feasible region. Thus, differentiating them with respect to
$p_{\mathrm{T}}$, and setting the derivatives to 0, we find the
optimal transmission probability $p_{\mathrm{T}}^{\ast}$ which
maximizes the expected number of successful receptions in each slot
as
\begin{equation}\label{eqn:optimal-tx-prob}
\begin{array}{lll}
    p_{\mathrm{T}}^{\ast} \;=\; \left\{ \begin{array}{cl}
                \displaystyle \frac{\tau^{\frac{2}{\eta}} +2 - \sqrt{(\tau^{\frac{2}{\eta}}-1)^2 + 3}}{3\tau^{\frac{2}{\eta}}}, \;& \ \mathrm{if} \ \ \tau <
                1  \vspace{0.2cm}\\
                \displaystyle \frac{\tau^{-\frac{2}{\eta}} -4 + \sqrt{(\tau^{-\frac{2}{\eta}}-1)^2 + 3}}{3(\tau^{-\frac{2}{\eta}} -2)}, \;& \ \mathrm{if} \ \ \tau \geq 1
                \end{array}\right.

\end{array}
\end{equation}

\end{example}

\begin{figure}[t]
\centering
\epsfig{file=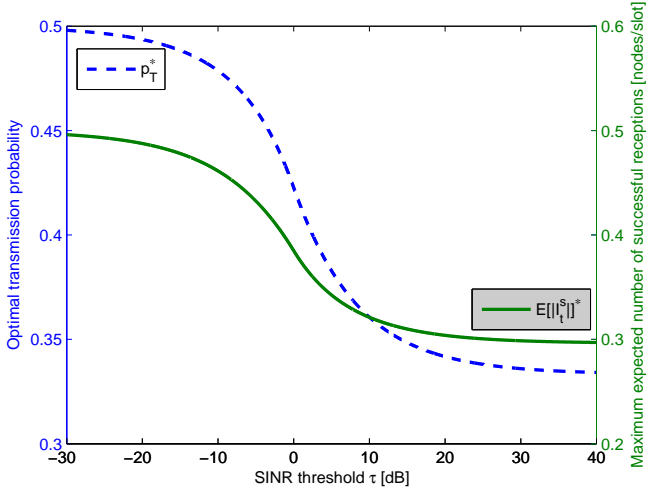,angle=0,width=0.5\textwidth}
\caption{Optimal transmission probability $p_{\mathrm{T}}^{\ast}$
and corresponding maximum expected number of successful receptions
$E[|\mathbf{I}_t^s|]^{\ast}$ for the three node example}
\label{fig:optimal-pt}
\vspace{0cm}
\end{figure}

In Fig. \ref{fig:optimal-pt}, we plot \eqref{eqn:optimal-tx-prob} as
a function of $\tau$ with path loss exponent $\eta=4$.
It can be seen that as threshold $\tau$ increases,
$p_{\mathrm{T}}^{\ast}$ decreases. This is because at a higher
threshold, it becomes more difficult for multiple transmitters to
simultaneously satisfy the SINR criterion. We can also see that the
maximum expected number of successful receptions
$E[|\mathbf{I}_t^s|]^{\ast}$ decreases as the threshold $\tau$
increases. However, it does not necessarily mean that we will have a
reduced number of successful receptions in a unit time, because the
transmission duration will also become shorter. We will look at this
issue in the subsequent section.
\begin{remark}
In \cite{borbash:asynchronous}, the optimal transmission probability
under the collision channel model was derived, and was given by the
inverse of the total number of nodes as in the original slotted
ALOHA system. Recall that as $\tau$ goes to $\infty$, our model
accommodates the collision channel model. In
(\ref{eqn:optimal-tx-prob}), we have $p_{\mathrm{T}}^{\ast}=1/3$ as
$\tau$ goes to $\infty$, which is consistent with the quoted result.
Therefore, we conclude that the optimal transmission probability
derived under collision channel model is fundamentally assuming the
\emph{worst prior} on the channel conditions by neglecting all the
possibilities of succeeding in the presence of interference.
\end{remark}

\subsection{On the Effect of the SINR Threshold}\label{sec:effect_of_the_threshold}

\begin{figure}[t]
\centering
\epsfig{file=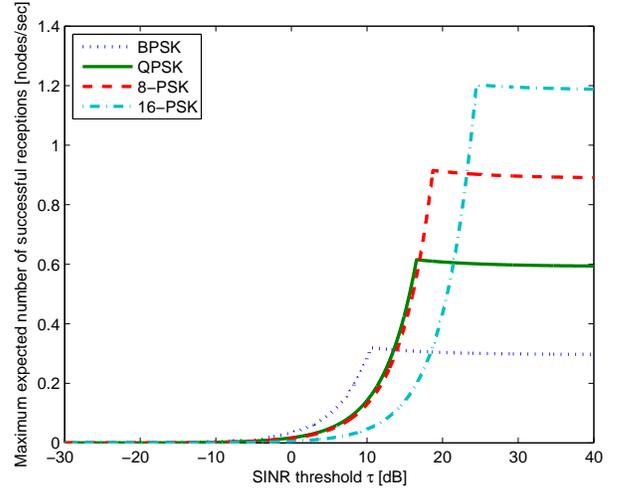,angle=0,width=0.5\textwidth}
\caption{The maximum expected number of successful receptions per
second under the $M$-PSK modulation scheme}
\label{fig:max_eh_per_sec}
\end{figure}

For a given modulation scheme and target BER, the data rate is an
increasing function of the SINR threshold $\tau$
\cite{goldsmith:wireless}. Hence, if the number of bits to be
transmitted is fixed and we increase $\tau$, the slot duration must
be shortened due to the increased rate. However, as we have seen in
the previous section, the transmission probability $p_{\mathrm{T}}$
needs to be lowered. To investigate this interrelationship, let us
consider the $M$-PSK modulation scheme\footnote{We could use the
Shannon capacity formula but it would not make practical sense since
it gives an asymptotic limit of the rate with arbitrarily small
probability of error and arbitrarily long block length.} in which
its symbol rate [symbols/sec/Hz] at a given target BER $z$ is given
by
\begin{equation*}\label{eqn:mpsk}
\begin{array}{lll}
    \displaystyle {R_s} \;\approx\; \left\{ \begin{array}{ll}
                \min{\left\{\frac{2 \tau}{\left[Q^{-1}(z)\right]^2}, R_s^{\mathrm{m}} \right\}}, \;& M=2 \mathrm{(BPSK)}\\
                \min{\left\{\frac{\tau}{2 \left[Q^{-1}(z)\right]^2}, R_s^{\mathrm{m}} \right\}}, \;& M=4 \mathrm{(QPSK)}\\
                \min{\left\{\frac{2 \tau \sin^2(\pi/M)}{\left[Q^{-1}\left(\frac{z \log_2 M}{2}\right)\right]^2}, R_s^{\mathrm{m}} \right\}}, \;&
                M=2^n, n>2
                \end{array}\right.

\end{array}
\end{equation*}
where $Q(z)$ is the probability that the standard normal random
variable is greater than $z$. The maximum symbol rate
$R_s^{\mathrm{m}}$ is given by $R_s^{\mathrm{m}}=1/k_g$, where $k_g$
is the constant that depends on the pulse shape of the analog
signal. Without loss of generality, we set $k_g = 1$ (i.e., the
raised cosine pulse with roll-off factor of 1). Denote by $W$ the
number of bits to be transmitted in each time slot. Then, the
transmission duration is given by
\begin{math}\label{eqn:slot-duration}
T_{\mathrm{slot}} = \frac{W}{R_s B \log_2 M}.
\end{math}
Consequently, we can redraw the maximum expected number of
successful receptions as in Fig. \ref{fig:max_eh_per_sec}, where we
set $\eta=4$, $z=10^{-6}$, $W=1$ bit, and $B=1$ Hz, respectively.
Note that in the figure, the units were changed from [nodes/slot] to
[nodes/sec].

\subsection{Performance over Multiple Slots}\label{sec:multslot}

Because as time goes on, the set of discovered neighbors in each
time slot will overlap with the previously discovered ones, it is
important to know the performance of the discovery algorithm over
multiple slots. We start with a simple scenario in which the set of
successful transmitters $\mathbf{I}_t^s$ is independent from slot to
slot. This is an artificial scenario because the set
$\mathbf{I}_t^s$ is correlated in the sense that a node closer to
the receiver has a better chance of being discovered in any slots.
Therefore, the scenario where $\mathbf{I}_t^s$ is correlated over
slots will also be studied later in this section.

Assuming that the set of successful transmitters $\mathbf{I}_t^s$ is
independent from slot to slot implies that each neighboring node has
equal probability of success. Therefore, for a given number of
successful receptions $h_t$ and the total number of neighbors $J$,
the probability that a certain neighboring node belongs to the set
$\mathbf{I}_t^s$ is given by $h_t/J$ and, over $D$ multiple slots,
the probability is obtained by
\begin{multline}\label{eqn:m-slot-uncorrelated-2}
\mathrm{Pr} \Set{ k \in \bigcup_{t=1}^{D} \mathbf{I}_t^s |\ \!\!|
\mathbf{I}_1^s | = h_1, \dots, | \mathbf{I}_D^s | = h_D } \\
= 1 - \prod_{t=1}^{D} \left(1 - \frac{h_t}{J}\right)
\end{multline}
We call (\ref{eqn:m-slot-uncorrelated-2}) a slot-basis prediction to
distinguish from the Bernoulli approximation which will be given in
the sequel.

Because the event that a particular neighboring node belongs to the
set $\mathbf{I}_t^s$ can be approximated by the Bernoulli trial with
success probability $E[|\mathbf{I}_t^s |]/J$, the number of
discovering the particular node over $D$ multiple slots is a
binomial random variable with success probability $E[|\mathbf{I}_t^s
|]/J$ and the total number of trials $D$. For large $D$ and small
$E[|\mathbf{I}_t^s |]/J$, it can be further approximated by the
Poisson random variable with parameter $DE[|\mathbf{I}_t^s |]/J$
\cite[p. 435]{leon:probability}. Hence, the probability that the
particular node has been discovered over $D$ multiple slots is
approximately $1-\exp(-{DE[|\mathbf{I}_t^s |]}/{J})$, which is equal
to the probability that the Poisson random variable is non-zero.
Note that the slot-basis prediction in
(\ref{eqn:m-slot-uncorrelated-2}) and the above approximation do not
depend on the particular node index because we assumed that
$\mathbf{I}_t^s$ is independent from slot to slot. Therefore, it can
be viewed as the predicted fraction of neighbors discovered up to
time slot $D$.
\begin{figure}[t]
\centering
\epsfig{file=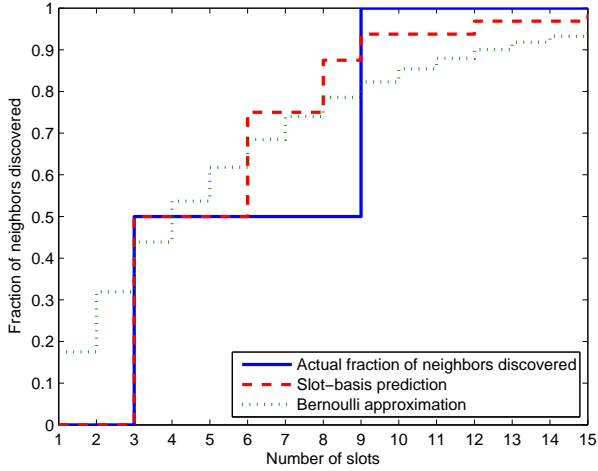,angle=0,width=0.5\textwidth}
\caption{The fraction of neighbors discovered by node $A$ for the
example transmit/listen pattern in Table \ref{table:txrxpattern1}}
\label{fig:discovered_fraction}
\end{figure}
\setlength{\tabcolsep}{3pt}
\begin{table}[t]
\caption{An example transmit/listen pattern of three nodes (the
letters T and L stand for transmit and listen,
respectively.)}\label{table:txrxpattern1} \centering
\begin{tabular}{c||ccccccccccccccc}\hline
Time slot & 1 & 2 & 3 & 4 & 5 & 6 & 7 & 8 & 9 & 10 & 11 & 12 & 13 &
14 & 15\\ \hline \hline

Node $A$ & T & T & L & L & T & L & T & L & L & T & T & L & T & T &
L\\ \hline

Node $B$ & L & L & T & L & T & T & L & T & L & L & L & L & T & T &
T\\ \hline

Node $C$ & L & T & T & L & T & L & L & T & T & T & T & T & T & T &
L\\ \hline
\end{tabular}
\vspace{-0.2cm}
\end{table}
In Fig. \ref{fig:discovered_fraction}, we plot the actual/predicted
fraction of neighbors discovered for the example transmit/listen
pattern in Table \ref{table:txrxpattern1}. For the figure, the SINR
threshold $\tau$ is set to 1 (in linear scale), and the transmission
probability $p_{\mathrm{T}}$ is set to 0.4226 which is the optimal
value by \eqref{eqn:optimal-tx-prob} with path loss exponent
$\eta=4$. Note that this is the optimal transmission probability
which maximizes the expected number of successful receptions at that
threshold value. Consequently, we obtain the expected number of
successful receptions per slot as 0.3849 by
(\ref{eqn:num-success-reception-n2-tau2}), and this value was used
to plot the Bernoulli approximation.

In practice, the probability that a particular node belongs to the
set of successful transmitters $\mathbf{I}_t^s$ depends on the
distance between the node and the receiver and, thus,
$\mathbf{I}_t^s$ is correlated over time slots. Therefore, we first
obtain the conditional probability that a particular node at the
specific distance from the receiver belongs to the set
$\mathbf{I}_t^s$ given the total number of successful receptions
$h_t$ as (see Appendix \ref{appendix:derivation_of_correlated_case})
\begin{multline}\label{eqn:m-slot-correlated-1}
\mathrm{Pr} \Set{ k \in \mathbf{I}_t^s |\ \!\!
|\mathbf{I}_t^s|=h_t, r_k=r'} \\
= \frac{\sum_{n=h_t}^{J} { J-1 \choose n-1} {p_{\mathrm{T}}}^{n} (1
- p_{\mathrm{T}})^{J - n} \zeta_n }{\sum_{n=h_t}^{J} {J \choose n}
{p_{\mathrm{T}}}^n (1-p_{\mathrm{T}})^{J-n} \left( \frac{n}{J}
\left(\zeta_n + \gamma_n - \xi_n \right) + \xi_n \right)}
\end{multline}
where $\zeta_n$, $\gamma_n$, and $\xi_n$ are defined as
\begin{align}
\zeta_n & = { n-1 \choose h_t-1} f_{1,n} f_{2,n}^{h_t-1}
(1-f_{2,n})^{n-h_t}\label{eqn:correlated-zeta}\\
\gamma_n & = {n-1 \choose h_t} f_{2,n}^{h_t} (1-f_{2,n})^{n-h_t-1}
(1-f_{1,n})\label{eqn:correlated-gamma}\\
\xi_n & = { n \choose h_t} f_n^{h_t}
(1-f_n)^{n-h_t}\label{eqn:correlated-xi}
\end{align}
and $f_{1,n}$ is the probability that a transmitter located at
distance $r'$ from the receiver will succeed among $n(\geq1)$
simultaneous transmissions and $f_{2,n}$ is the probability that a
transmitter at an arbitrary distance will succeed among $n(>1)$
simultaneous transmissions given that one of the other transmitter
is located at distance $r'$ from the receiver. The expressions for
$f_{1,n}$ and $f_{2,n}$ are given in Appendix
\ref{appendix:derivation_of_correlated_case}. The function $f_n$ is
the shorthand notation for (\ref{eqn:capture-cal}). Using the
conditional probability in \eqref{eqn:m-slot-correlated-1}, the
probability that a particular node at distance $r'$ has been
discovered over multiple slots can be obtained similarly with
\eqref{eqn:m-slot-uncorrelated-2}. To get a better understanding on
the conditional probability $\eqref{eqn:m-slot-correlated-1}$, let
us consider the following example which is in line with the Example
\ref{example:opt_tx_prob}.

\begin{example}
Since each node has two neighbors, it is obvious that $\mathrm{Pr}
\Set{ k \in \mathbf{I}_t^s |\ \!\! |\mathbf{I}_t^s|=0, r_k=r'} = 0$
and $\mathrm{Pr} \Set{ k \in \mathbf{I}_t^s |\ \!\!
|\mathbf{I}_t^s|=2, r_k=r'} = 1$. For $h_t=1$, after some
calculation, we obtain
\begin{multline}\label{eqn:m-slot-correlated-example-result-fianl}
\displaystyle \mathrm{Pr} \Set{ k \in \mathbf{I}_t^s |\ \!\!
|\mathbf{I}_t^s|=1, r_k=r'}\\
= \displaystyle \frac{1-p_{\mathrm{T}} +
p_{\mathrm{T}}\left(1-{r'}^2\right)^2 } {2(1-p_{\mathrm{T}}) +
\left( \left(1-{r'}^2 \right)^2 + {r'}^4 \right) p_{\mathrm{T}}}
\end{multline}
where the discovery range $R_0$ in \eqref{eqn:distribution-fn} is
normalized to 1 and the threshold $\tau$ is set to 1.
\begin{figure}[t]
\centering
\epsfig{file=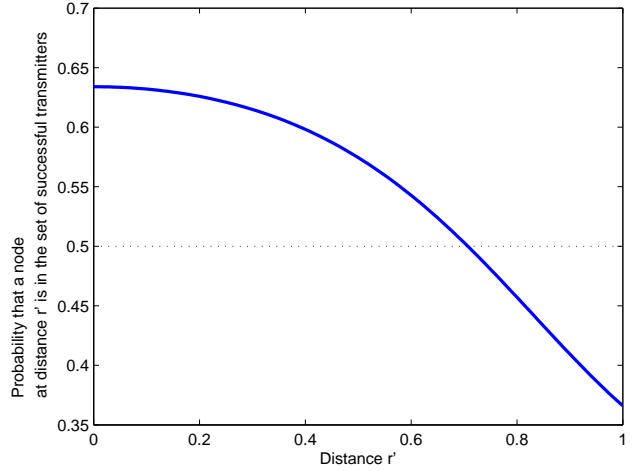,angle=0,width=0.5\textwidth}
\caption{The probability that a node at distance $r'$ from the
reference node is in the set of successful transmitters for the
three node example given the number of successful transmissions
$h_t=1$} \label{fig:correlated-oneslot}
\end{figure}
We plot the result in Fig. \ref{fig:correlated-oneslot} where all
other parameters are set identical with those used for Fig
\ref{fig:discovered_fraction}. Note that even if $r'=0$, the
probability is not equal to 1 because of the randomness on node's
transmission.
\end{example}

\section{Physical Layer Signal Processing: Detection of the Transmitting Neighbors}\label{sec:matched_filter}

\subsection{Classical Approach Using a Bank of Matched
Filters}

To decide the existence of a signal from a particular node, we
require the use of matched filters in which the outputs of the
filters are compared to a certain threshold. Note that such a
decision is subject to probabilistic errors such as the \emph{false
alarm} and \emph{miss} and, thus, the threshold needs to be chosen
in some optimum way. Since the set of transmitting neighbors and
their signal amplitudes are all unknown and hard to be tracked, we
simplify it by assuming that the sum of interfering signals and the
noise act as another noise process
$\mathbf{n}_t'=\{n_{t,1}',...,n_{t,L}'\}^T$ whose samples are i.i.d.
with $\mathcal{N}(0, N')$. The average noise power $N'$ can be
computed as
\begin{equation}\label{eqn:noise_power_new}
N' = N + \sum_{n=1}^{\bar{J}-1} {\bar{J}-1 \choose n}
{p_{\mathrm{T}}}^n (1-p_{\mathrm{T}})^{\bar{J}-1-n} n \bar{P}_{rx}
\end{equation}
where $\bar{J}$ and $\bar{P}_{rx}$ is the average number of
neighbors and the received signal power, respectively. Note that
this is a reasonable approximation if there is a large number of
nodes, and it is commonly applied to the analysis of cellular code
division multiple access (CDMA) systems \cite{goldsmith:wireless}.

\begin{figure}[t]
\centering
\epsfig{file=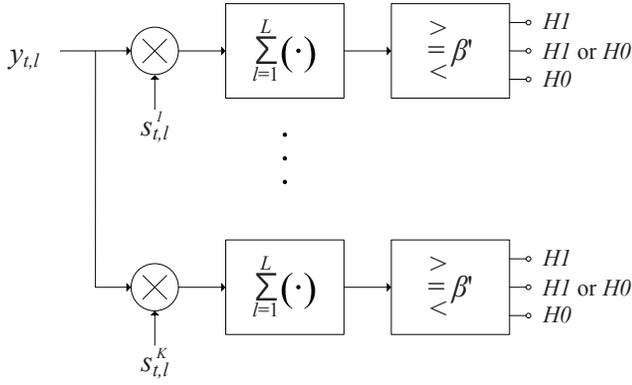,angle=0,width=0.46\textwidth}
\caption{Bank of matched filters} \label{fig:matched_filter}
\vspace{0cm}
\end{figure}

Based on the above approximation, we formulate the binary hypothesis
testing problem for the transmission of a particular node $k$ as
\begin{equation}\label{eqn:mf_hypothesis_problem_new}
\begin{array}{rl}
& \displaystyle{H_0:\mathbf{y}_t = \mathbf{n}_t'}\\
\textrm{versus}&\\
& \displaystyle{H_1:\mathbf{y}_t ={g_t^{k} \mathbf{s}_t^{k}} +
\mathbf{n}_t'}
\end{array}
\end{equation}
which is a \textit{composite} hypothesis testing because the signal
$\mathbf{s}_t^k$ is known but still not its amplitude. For the
problem, a \textit{generalized likelihood ratio test} method would
provide a simple decision rule as\footnote{Note that the decision
rule in (\ref{eqn:mf_glrt}) is optimal for $g_t^k$ near $0$ which
holds for most of the practical situations.} \cite[p.
51]{poor:introduction}
\begin{equation}\label{eqn:mf_glrt}
\displaystyle \frac{1}{N'} \sum_{l=1}^L y_{t,l} s_{t,l}^k \displaystyle{\begin{array}{c} \vspace{-0.1cm}{H_1}\\
\vspace{-0.03cm} \gtrless\\ {H_0} \end{array}} \beta
\end{equation}
This structure is depicted in Fig. \ref{fig:matched_filter}, where
$\beta' = \beta N'$.
Note that each node transmits with probability $p_{\mathrm{T}}$ and
listens with probability $1 - p_{\mathrm{T}}$, and they are the
\textit{prior} probabilities for each hypothesis.

Given the cost structure $C_{ij}$ which is the cost incurred by
choosing hypothesis $H_i$ when hypothesis $H_j$ is true, the optimum
threshold for the minimum average cost is given by
\begin{math}
\beta = \frac{\pi_0 (C_{10} - C_{00})}{\pi_1 (C_{01} - C_{11})},
\end{math}
where $\pi_i$ is the \textit{prior} probability of hypothesis $H_i$.
Under the \textit{minimum-probability-of-error} criterion where the
cost assignment is done by $C_{ij}=0$ for $i=j$, and $C_{ij}=1$ for
$i \neq j$, the threshold is obtained by $\beta = \pi_0 / \pi_1 = (1
- p_{\mathrm{T}})/ p_{\mathrm{T}}$.

\subsection{Random Set Theory-based Approach}\label{sec:nd_rst}

\subsubsection{Motivation and Background}


The classical approach using a bank of matched filters holds a
certain desired property; that is, its complexity does not scale
with the number of nodes by averaging out the effect of random
interferences. However, the performance would be worse than the
class of decorrelator detectors. On the other hand, total number of
transmitting neighbors and their entities are all random in our
problem setting and, thus, standard decorrelator detectors are not
directly applicable because they fundamentally assume a fixed number
of transmitters with known entities.
%
%
This problem naturally falls in the purview of RST which generalizes
standard probability theory by assigning \emph{sets}, rather than
values, to random outcomes (see Appendix \ref{appendix:rst} and
references therein).
%
%
RST has been applied before in the context of multi-source data
fusion and multi-target identification problems
\cite{goodman:mathematics, ali:random, vihola:random,
mahler:engineering, mahler:random, mahler:statistical} and,
recently, multi-user detection problem in a dynamic environment
\cite{biglieri:multiuser}. The utility of RST mostly comes from the
fact that we can readily treat the random behavior of a random
number of entities as a single random set having likelihood.

%

Mathematically, a random set $\mathbf{X}$ is defined as a mapping
from a sample space $\Omega$ to a power set
$\mathcal{P}(\mathbb{S})$ of a hybrid space $\mathbb{S}$. It is
referred to as a random \textit{finite} set, if for all $\omega \in
\Omega$, the set is finite (i.e., $|\mathbf{X}(\omega)|<\infty$).
The hybrid space $\mathbb{S}\triangleq \mathbb{R}^d \times U$ is the
Cartesian product of a $d$-dimensional Euclidean space
$\mathbb{R}^d$ and a finite discrete space $U$. To illustrate the
use of the RST-based method, we focus on the case where the hybrid
space $\mathbb{S}$ is given by a finite discrete space as
$\mathbb{S}=\{1,...,K\}$, i.e., the random set $\mathbf{X}_t$ is
simply equal to the \textit{unknown} set of transmitting neighbors
at that time slot, which was previously denoted by $\mathbf{I}_t$ in
\eqref{eqn:received-signal}. Additionally, if the amplitudes of the
signals from transmitting neighbors are of interest, we can defined
it as $\mathbb{S} = \{1,\dots, K\} \times \mathbb{R}_+$, where
$\mathbb{R}_+$ is the non-negative real space. In the following, we
estimate the random set $\mathbf{X}_t$ based on the observed signal
$\mathbf{y}_t$ in each time slot.

%

\subsubsection{Defining the Estimator}\label{sec:rst_estimator}

Since the number of transmitting neighbors in each time slot and the
total number of neighbors are connected, both the random set
$\mathbf{X}_t$ and the total number of neighbors $J$ need to be
jointly estimated as
\begin{equation*}\label{eqn:rst_estimator}
\arg \max_{\left({\mathbf{X}}_t', {J}' \right)} f_{\mathbf{X}_t,
J|\mathbf{Y}_t} \left( \mathbf{X}_t', J'|\mathbf{y}_t \right)
\end{equation*}
where $f_{\mathbf{X}_t, J|\mathbf{Y}_t}(\cdot)$ is the likelihood of
random set $\mathbf{X}_t$ and $J$ neighbors given the received
signal $\mathbf{Y}_t = \mathbf{y}_t$. By Bayes rule,
$f_{\mathbf{X}_t, J|\mathbf{Y}_t} \left( \mathbf{X}_t',
J'|\mathbf{y}_t \right)$ is proportional to
\begin{equation*}\label{eqn:rst_likelihood_bayes}
\begin{array}{l}
\displaystyle {f_{\mathbf{Y}_t|\mathbf{X}_t, J} \left( \mathbf{y}_t
| \mathbf{X}_t', J'\right) f_{\mathbf{X}_t|J}(\mathbf{X}_t'| J')
f_{J}(J')}
\end{array}
\end{equation*}
where $f_{\mathbf{Y}_t|\mathbf{X}_t, J} (\cdot)$ is the likelihood
of received signal $\mathbf{Y}_t$ given $\mathbf{X}_t =
\mathbf{X}_t'$ and $J=J'$, $f_{\mathbf{X}_t|J} (\cdot)$ is the
likelihood of random set $\mathbf{X}_t$ given $J=J'$, and
$f_{J}(\cdot)$ is the likelihood of having $J$ neighbors.
Consequently, the joint MAP estimator of the random set
$\mathbf{X}_t$ and the total number of neighbors $J$ is obtained by
\begin{equation}\label{eqn:rst_estimator_map}
\arg \max_{\left({\mathbf{X}}_t', J' \right)}
{f_{\mathbf{Y}_t|\mathbf{X}_t, J} \left( \mathbf{y}_t |
\mathbf{X}_t', J'\right) f_{\mathbf{X}_t|J}(\mathbf{X}_t'| J')
f_{J}(J')}
\end{equation}
In order to run the above estimator, it is required to specify the
densities $f_{\mathbf{Y}_t|\mathbf{X}_t, J} (\cdot)$,
$f_{\mathbf{X}_t|J} (\cdot)$, and $f_{J}(\cdot)$. We outline how
these densities are obtained.
%

\subsubsection{Specifying the Densities}\label{sec:rst_densities}


The received signal $\mathbf{y}_t$ not only depends on the set of
transmitting neighbors $\mathbf{X}_t$, but also depends on their
signal amplitudes $\mathbf{g}_t = \{g_t^k, \forall k \in
\mathbf{X}_t\}$. Since the amplitudes are all random, we take
average by assuming that the amplitudes of the nodes are independent
of each other as 
%
\begin{align}\label{eqn:rst_case1_density1_1}
\displaystyle & f_{\mathbf{Y}_t|\mathbf{X}_t, J}
(\mathbf{y}_t|\mathbf{X}_t', J') \nonumber \\
& = \displaystyle \! \int_0^{\infty} \!\!\! \cdots \int_0^{\infty}
\!\!\! f_{\mathbf{Y}_t|\mathbf{X}_t, \mathbf{g}_t, J}
(\mathbf{y}_t|\mathbf{X}_t', \mathbf{g}_t', J') dF_g(g_1') \cdots
dF_g(g_{|\mathbf{X}_t'|}')
\end{align}
where $f_{\mathbf{Y}_t|\mathbf{X}_t, \mathbf{g}_t,
J}(\mathbf{y}_t|\mathbf{X}_t', \mathbf{g}_t', J')$ is the likelihood
of received signal $\mathbf{Y}_t$ given
$\mathbf{X}_t=\mathbf{X}_t'$, $\mathbf{g}_t = \mathbf{g}_t'$, and
$J=J'$. The set $\mathbf{g}_t'=\{g_1', \dots,
g_{|\mathbf{X}_t'|}'\}$ and $F_g(\cdot)$ denote the realization of
$\mathbf{g}_t$ and the common cumulative distribution of the
received signal amplitude, respectively. 
Note that by further conditioning the received signal $\mathbf{y}_t$
on the set of transmitting neighbors and their signal amplitudes,
the only randomness remaining is in the noise $\mathbf{n}_t$.
Therefore, it is given by
%
%
\begin{align*}
& f_{\mathbf{Y}_t|\mathbf{X}_t, \mathbf{g}_t, J}
(\mathbf{y}_t|\mathbf{X}_t', \mathbf{g}_t',
J') \nonumber\\
& = \frac{1}{\left({2 \pi N}\right)^{L/2}} \exp\left({-\frac{1}{2N}
\sum_{l=0}^L \left(y_{t,l} - \sum_{k \in \mathbf{X}_t'} g_{m(k)}'
s_{t,l}^k \right)^2} \right)
\end{align*}
where the bijective function $m(k)$ has been introduced to map the
elements in $\mathbf{X}_t'$ to the elements in the set $\{1,\dots,
|\mathbf{X}_t'|\}$. For example, if $\mathbf{X}_t' = \{2,5\}$, then
$m(2)=1$ and $m(5)=2$.



In order to obtain $f_{\mathbf{X}_t|J} (\mathbf{X}_t'|J')$, we first
obtain the \textit{belief mass} of a random set $\mathbf{X}_t$ for a
given number of neighbors $J=J'$ as (see Appendix
\ref{appendix:rst})
%
%
\begin{align*}
\beta_{\mathbf{X}_t|J}(\mathbf{C}|J') = \displaystyle
\sum_{n=0}^{J'} \sum_{\mathbf{B}: \mathbf{B} \subseteq \mathbf{C},
|\mathbf{B}|=n} \mathrm{Pr}\{\mathbf{X}_t=\mathbf{B}|J=J'\}
\end{align*}
%
where $\mathbf{C}$ is a closed subset of the space $\mathbb{S}$, and
$\mathbf{B}$ is a realization of the random set $\mathbf{X}_t$. Let
us first derive $\beta_{\mathbf{X}_t|J}(\mathbf{C}|J')$ for a
particular example as follows.
\begin{example}\label{example_belief_mass}
Set $\mathbb{S}=\{1,2,3\}$, $\mathbf{C}=\mathbb{S}$, and $J'=2$. 1)
The set $\{\mathbf{B}: \mathbf{B} \subseteq \mathbf{C},
|\mathbf{B}|=0\}$ is given by $\{\emptyset\}$ and, thus,
$\sum_{\mathbf{B}: \mathbf{B} \subseteq \mathbf{C}, |\mathbf{B}|=0}
\mathrm{Pr}\{\mathbf{X}_t=\mathbf{B}|J=2\}=\left(1- p_{\mathrm{T}}
\right)^2$. 2) The set $\{\mathbf{B}: \mathbf{B} \subseteq
\mathbf{C}, |\mathbf{B}|=1\}$ is given by $\{ \{1\}, \{2\}, \{3\}
\}$ and, thus, $\sum_{\mathbf{B}: \mathbf{B} \subseteq \mathbf{C},
|\mathbf{B}|=1} \mathrm{Pr}\{\mathbf{X}_t=\mathbf{B}|J=2\}=3
p_{\mathrm{T}} \left(1- p_{\mathrm{T}} \right)$. 3) The set
$\{\mathbf{B}: \mathbf{B} \subseteq \mathbf{C}, |\mathbf{B}|=2\}$ is
given by $\{ \{1,2\}, \{1,3\},$ $ \{2,3\} \}$ and, thus,
$\sum_{\mathbf{B}: \mathbf{B} \subseteq \mathbf{C}, |\mathbf{B}|=2}
\mathrm{Pr}\{\mathbf{X}_t=\mathbf{B}|J=2\}=3 {p_{\mathrm{T}}}^2$. 4)
Since $J'=2$, the probability that a random set $\mathbf{X}_t$ is
equal to $\mathbf{B}$ for $|\mathbf{B}|>2$ is zero. Summing over all
possible $\mathbf{B}$'s yields $\beta_{\mathbf{X}_t|J}
(\mathbf{C}|J')=\sum_{n=0}^{2} {3 \choose n} {p_{\mathrm{T}}}^n
\left(1- p_{\mathrm{T}} \right)^{2-n}$.
\end{example}
For general cases, we have
\begin{equation*}\label{eqn:rst_case1_beliefmass_definition}
\begin{array}{l}
\beta_{\mathbf{X}_t|J}(\mathbf{C}|J') = \displaystyle
\sum_{n=0}^{J'} {|\mathbf{C}| \choose n} {p_{\mathrm{T}}}^n \left(1-
p_{\mathrm{T}} \right)^{J'-n}
\end{array}
\end{equation*}
The \textit{belief density} $f_{\mathbf{X}_t|J} (\mathbf{X}_t'|J')$
is obtained by taking the \textit{set derivative} of the
\textit{belief mass} obtained above. For the case where the hybrid
space $\mathbb{S}$ is comprised only of the discrete space, it can
be readily obtained through the following M\"{o}bius inversion
formula as (see Appendix \ref{appendix:rst})
\begin{equation*}\label{eqn:rst_case1_mobius}
f_{\mathbf{X}_t|J}(\mathbf{X}_t'|J') = \sum_{\mathbf{C}\subseteq
\mathbf{X}_t'} (-1)^{|\mathbf{X}_t' \setminus \mathbf{C}|}
\beta_{\mathbf{X}_t|J}(\mathbf{C}|J')
\end{equation*}
\begin{example}\label{example_belief_density}
Take $\mathbf{X}_t'=\{1,3\}$ and $J'=2$, then the set
$\{\mathbf{C}:\mathbf{C}\subseteq \mathbf{X}_t'\}$ is given by
$\{\emptyset, \{1\}, \{3\}, \{1,3\} \}$. 1) For
$\mathbf{C}=\emptyset$, $\beta_{\mathbf{X}_t|J}(\mathbf{C}|J') =
\left(1- p_{\mathrm{T}} \right)^{2}$. 2) For $\mathbf{C}=\{1\}$,
$\beta_{\mathbf{X}_t|J}(\mathbf{C}|J') = \left(1- p_{\mathrm{T}}
\right)^{2} + p_{\mathrm{T}}\left(1- p_{\mathrm{T}} \right)$. 3) For
$\mathbf{C}=\{3\}$, $\beta_{\mathbf{X}_t|J}(\mathbf{C}|J') =
\left(1- p_{\mathrm{T}} \right)^{2} + p_{\mathrm{T}}\left(1-
p_{\mathrm{T}} \right)$. 4) For $\mathbf{C}=\{1,3\}$,
$\beta_{\mathbf{X}_t|J}(\mathbf{C}|J') = \left(1- p_{\mathrm{T}}
\right)^{2} + 2p_{\mathrm{T}}\left(1- p_{\mathrm{T}} \right) +
{p_{\mathrm{T}}}^2$. Summing over all the possible $\mathbf{C}$'s by
considering the sign of the terms yields
$f_{\mathbf{X}_t|J}(\mathbf{X}_t'|J')={p_{\mathrm{T}}}^2$.
\end{example}
Similarly, for general cases, we have
\begin{equation*}\label{eqn:rst_case1_beliefdensity}
\begin{array}{ll}
    f_{\mathbf{X}_t|J}(\mathbf{X}_t'|J') \! \;=\; \! \left\{ \begin{array}{cl}
                \displaystyle \!\! {p_{\mathrm{T}}}^{|\mathbf{X}_t'|} \left(1- p_{\mathrm{T}}
\right)^{J' - |\mathbf{X}_t'|}, \;& \!\! \textrm{for} \ |\mathbf{X}_t'| \leq J'\\
                \displaystyle \!\! 0, \;& \!\! \textrm{otherwise}
                \end{array}\right.

\end{array}
\end{equation*}
Note that the scenario in which $\mathbf{X}_t$ contains only the
identity of the transmitters is the simplest case that can be solved
by standard probability theory as well. The usefulness of RST comes
when we extend the set $\mathbf{X}_t$ so that additional parameters,
such as signal amplitudes, can be estimated at the same time.


The density $f_{J}(J')$ is the probability that there are $J'$
number of neighbors. By defining the discovery region as the circle
of radius $R_0$, from Section \ref{sec:system}, the number of nodes
inside the discovery region follows a Poisson random variable with
parameter $\lambda \pi {R_0}^2$. Notice that there is no definite
way of choosing $R_0$ because the decisions on whether a particular
node is my neighbor are inconclusive due to the continuity of signal
strength together with the random effect of noise and fading.
However, as will be shown in the numerical example, the estimator
gives more weights to an appropriate size of the set which is likely
to be occurred at that size of discovery region.

\section{Numerical Results}\label{sec:numerical_results}

\begin{figure*}[t]
\centering \subfigure[Discretization of the distance
]{\label{fig:rst_numerical_example_sub_a}\epsfig{file=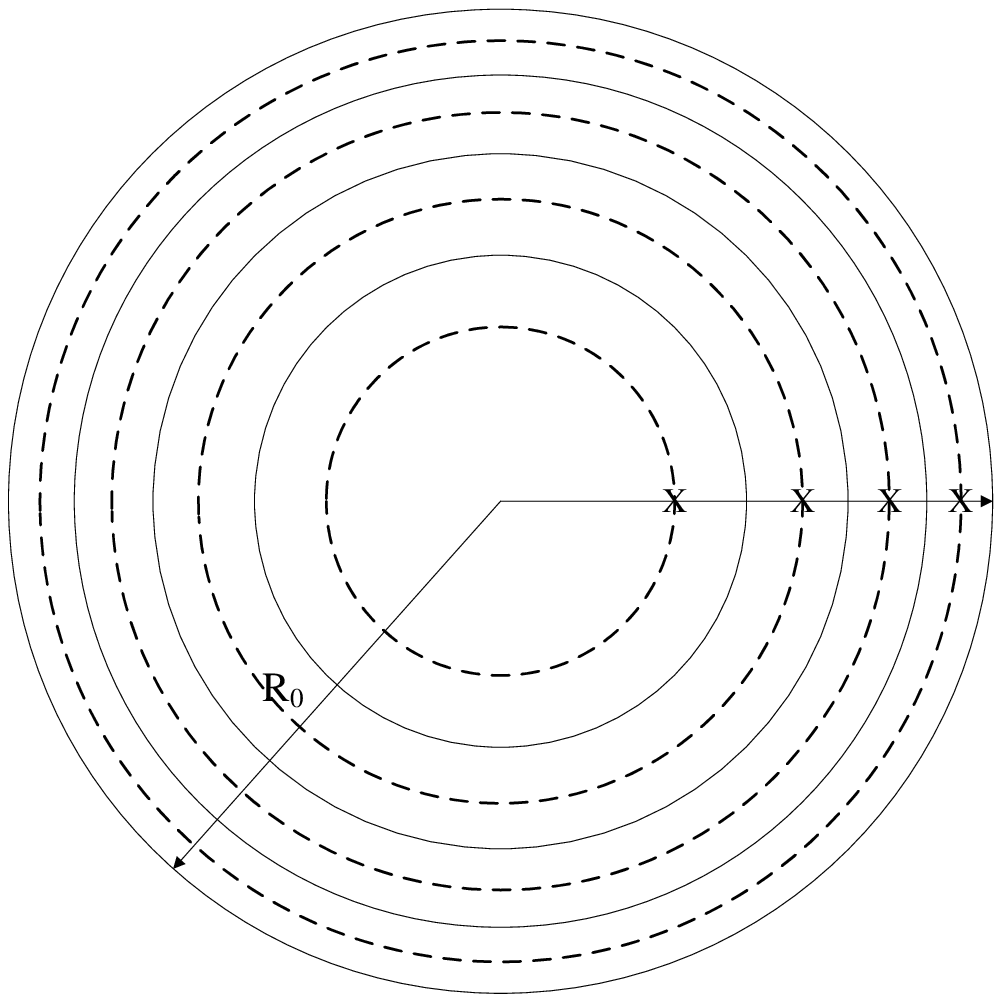,angle=0,width=0.29\textwidth}}
\hspace{0.4cm}\centering \subfigure[Deployment scenario 1
]{\label{fig:rst_numerical_example_sub_b}\epsfig{file=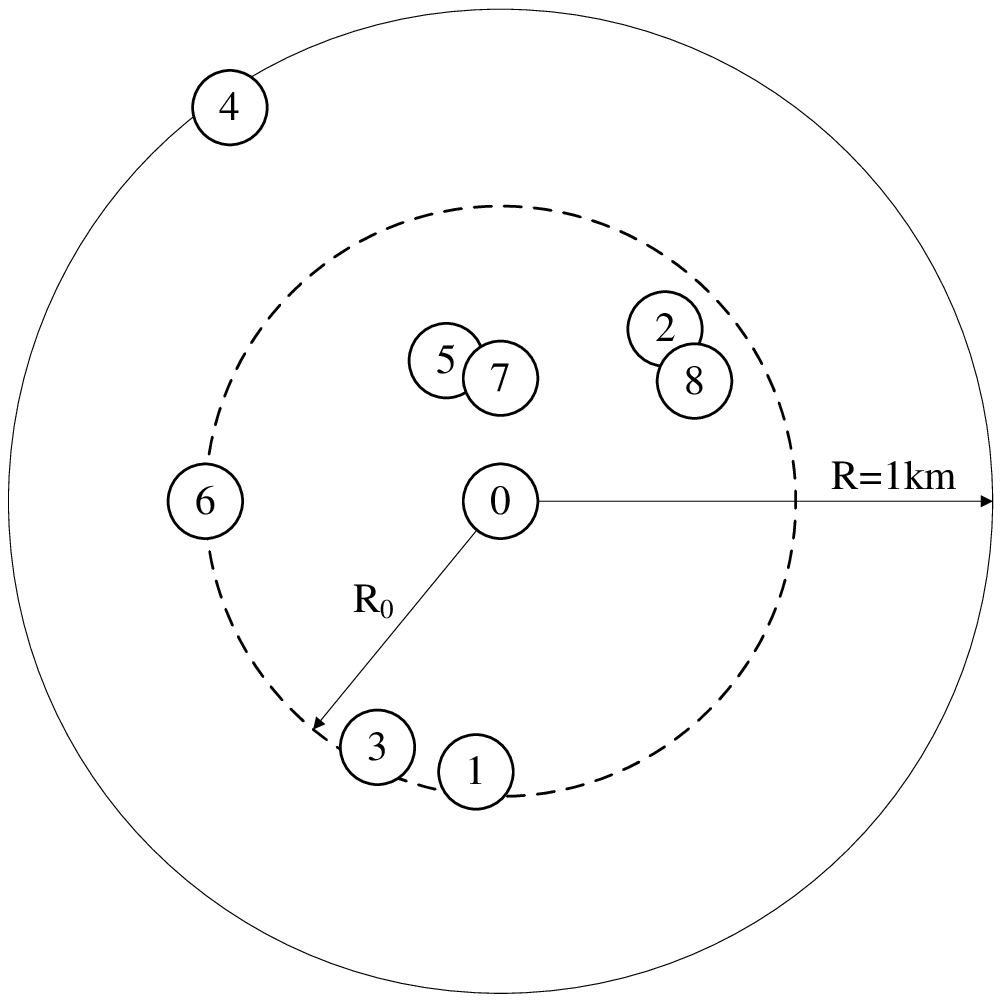,angle=0,width=0.29\textwidth}}
\hspace{0.4cm}\centering \subfigure[Deployment scenario 2]
{\label{fig:rst_numerical_example_sub_c}\epsfig{file=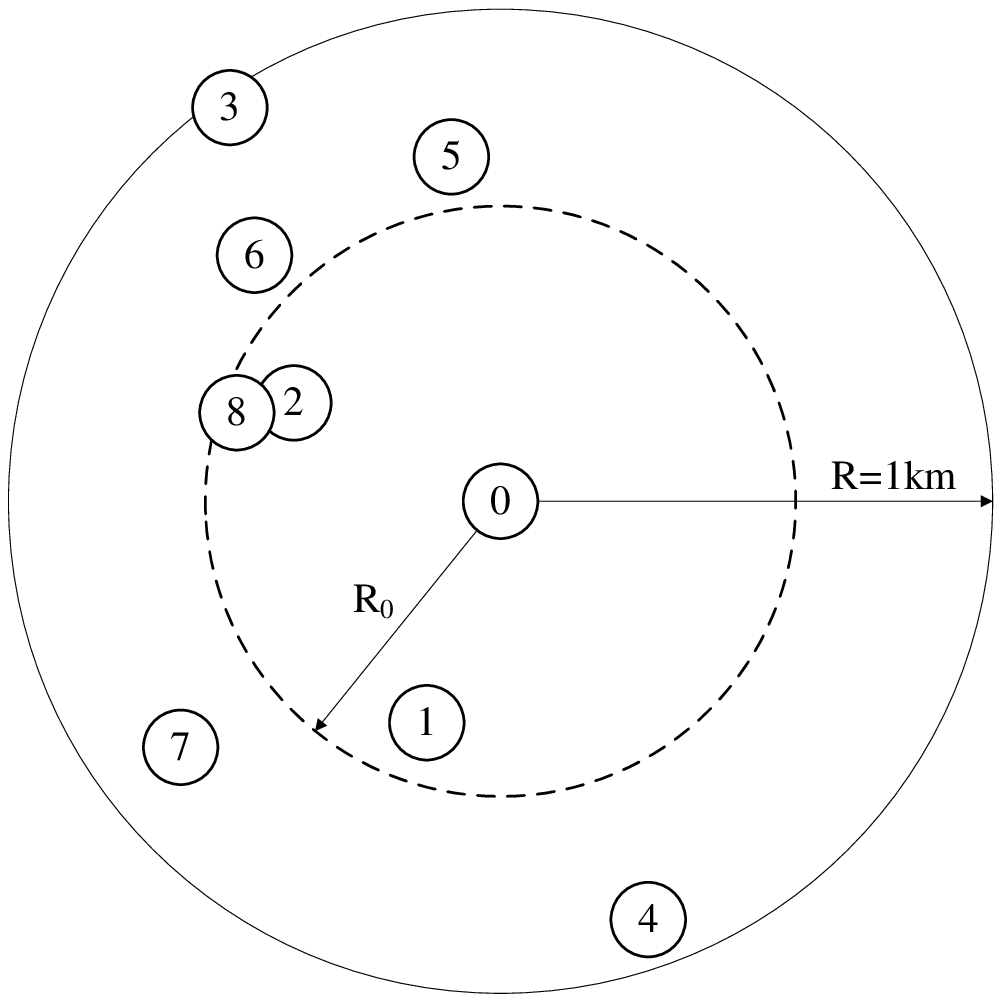,angle=0,width=0.29\textwidth}}
\caption{Discretization of the distance for fast computation of
(\ref{eqn:rst_case1_density1_1}) and example deployment scenarios}
\label{fig:rst_numerical_example}
\end{figure*}

Since comparing the performance of the discovery algorithm under
collision channel to that under multipacket reception channel is
trivial and obvious, we focus on the comparison of the physical
layer signal processing methods for the detection of transmitting
neighbors. For the simulation, a total of 8 wireless sensor nodes
are uniformly deployed over the region of interest which is modeled
as a circle of radius $R$ which is set to 1 km as shown in Fig.
\ref{fig:rst_numerical_example}. The reference node denoted by index
0 is assumed to be located at the center of the circle. For the
wireless channel, we consider a simple path loss model which only
depends on the transmitter-receiver distance, and the path loss
exponent $\eta$ is set to 4. The transmission power $G$ is set to
-24 dBm\footnote{For example, the transmission power of the MICAz
Mote, a commercial wireless sensor node by Crossbow Technology,
Inc., is programmable in 8 steps from -24 to 0 dBm
\cite{xbow:micaz}.}, and the noise spectral density $N_0$ is set to
-173 dBm/Hz as commonly done. The bandwidth $B$ is set to 100 Hz
(which is artificially low but really irrelevant for our purposes
here). The transmission probability $p_{\mathrm{T}}$ is set to 0.5
(to better observe the effect of the multiple access). For the
signatures of the nodes, we used length 15 maximal-length sequences
\cite{stallings:wireless}.
%
%
Gold and Kasami sequences which have better correlation property can
be used, but comparing the performance of different codes is beyond
the scope of this paper.

%
%
%
%
%
%
%
%
%


Note that the computation of $f_{\mathbf{Y}_t|\mathbf{X}_t, J}
(\cdot)$ in \eqref{eqn:rst_case1_density1_1} is tedious because of
the multiple integrals. Hence, we transform the integrals to a
finite summation by discretizing the distance. To do this, we first
divide the discovery region of the reference node into a finite
number of strips having the same area and, after that, each strip is
further divided into two having the same area as shown in Fig.
\ref{fig:rst_numerical_example_sub_a}. By doing so, the probability
that a node is on either one of those radii becomes uniform. For the
numerical examples, we used 7 discrete points and, obviously, the
accuracy will improve as the number of points increases.
%
%
Also note that, since a total of 8 nodes are uniformly deployed over
the specified region of interest, the density $f_{J}(\cdot)$ on the
number of nodes inside the discovery region of radius $R_0$ can be
more accurately described by the binomial distribution $B(8,
\frac{{R_0}^2}{{R}^2})$, rather than the Poisson approximation. For
the classical approach, it is computed by $\beta'=\beta N'$ where
$\beta = (1-p_{\mathrm{T}})/p_{\mathrm{T}}=1$ and $N' = N_0B + 3.5
\bar{P}_{rx}$ by \eqref{eqn:noise_power_new} with $\bar{P}_{rx} =
3.3333\times10^{-7}G$ using the cumulative distribution function of
the received signal power in \eqref{eqn:cdf_rx_power_1} without
fading.

%
%
%
%
%
%
%
%
%

%
\fboxsep=0.4pt
\fboxrule = 0.25pt
%
%

%
\setlength{\tabcolsep}{1.4pt}
\begin{table}[t]
\caption{Detection results for deployment scenario 1 in Fig.
\ref{fig:rst_numerical_example_sub_b} (the letters H, C, F, and M
stand for \textit{hit}, \textit{correct rejection}, \textit{false
alarm}, and \textit{miss},
respectively.)}\label{table:detection_1}\centering
\vspace{0.1cm}
\subtable[RST-based approach ($R_0 = 1 \ \mathrm{km}$)]{
   \label{table:detection_1_rst_1}
\begin{tabular}{c||cccccccccccccccccccc}\hline
Time slot & 1 & 2 & 3 & 4 & 5 & 6 & 7 & 8 & 9 & 10 & 11 & 12 & 13 &
14 & 15 & 16 & 17 & 18 & 19 & 20\\ \hline\hline

Node 1 & C & H & C & C & C & C & H & C & H & C & H & C & H & C & C &
H & H & C & C & H
\\ \hline

Node 2 & H & H & H & H & H & C & C & C & H & C & C & H & C & C & C &
C & H & H & C & H
\\ \hline

Node 3 & H & H & C & C & C & C & C & H & H & \fbox{\textbf{F}} & H &
H & H & H & C & C & C & C & C & C
\\ \hline

Node 4 & C & C & C & \fbox{\textbf{F}} & H & C & H & H & C & C & C &
C & H & C & C & H& \fbox{\textbf{M}} & C & C & H
\\ \hline

Node 5 & H & C & H & H & H & C & C & C & C & C & C & H & C & C & H &
C & C & H & C & H
\\ \hline

Node 6 & H & C & C & H & H & C & C & C & H & C & C & C & H & H & C &
H & H & H & H & C
\\ \hline

Node 7 & H & H & C & C & H & C & H & C & C & C & H & C & H &
\fbox{\textbf{F}} & C & C & H & H & H & C
\\ \hline

Node 8 & C & H & H & H & H & H & H & H & C & C & H & H & H & C & H &
C & H & H & H & H
\\ \hline

\end{tabular}}
\vspace{0.3cm}
\setlength{\tabcolsep}{1.33pt}
\subtable[RST-based approach ($R_0 = 0.5 \ \mathrm{km}$)]{
   \label{table:detection_1_rst_2}
\begin{tabular}{c||cccccccccccccccccccc}\hline
Time slot & 1 & 2 & 3 & 4 & 5 & 6 & 7 & 8 & 9 & 10 & 11 & 12 & 13 &
14 & 15 & 16 & 17 & 18 & 19 & 20 \\ \hline\hline

Node 1 & C & H & C & C & C & C & H & C & H & C & H & C & H & C & C &
H & H & C & C & H
\\ \hline

Node 2 & H & H & H & H & H & C & C & C & H & C & C & H & C & C & C &
C & H & H & C & H
\\ \hline

Node 3 & H & H & C & C & C & C & C & H & H & \fbox{\textbf{F}} & H &
H & H & H & C & C & C & C & C & C
\\ \hline

Node 4 & C & C & C & C & \fbox{\textbf{M}} & C & H & H & C & C & C &
C & H & C & C & H & \fbox{\textbf{M}} & C & C & H
\\ \hline

Node 5 & H & C & H & H & H & C & C & C & C & C & C & H & C & C & H &
C & C & H & C & H
\\ \hline

Node 6 & H & C & C & H & H & C & C & C & H & C & C & C & H & H & C &
H & H & H & H & C
\\ \hline

Node 7 & H & H & C & C & H & C & H & C & C & C & H & C & H & C & C &
C & H & H & H & C
\\ \hline

Node 8 & C & H & H & H & H & H & H & H & C & C & H & H & H & C & H &
C & H & H & H & H
\\ \hline

\end{tabular}}
\vspace{0.3cm}
\setlength{\tabcolsep}{1.2pt}
\subtable[Classical approach using a bank of matched filters]{
   \label{table:detection_1_mf}
\begin{tabular}{c||cccccccccccccccccccc}\hline
Time slot & 1 & 2 & 3 & 4 & 5 & 6 & 7 & 8 & 9 & 10 & 11 & 12 & 13 &
14 & 15 & 16 & 17 & 18 & 19 & 20 \\ \hline\hline

Node 1 & C & H & C & C & C & \fbox{\textbf{F}} & H & C & H &
\fbox{\textbf{F}} & H & C & H & C & C & H & H & C & C & H
\\ \hline

Node 2 & H & H & H & H & H & C & C & C & H & \fbox{\textbf{F}} & C &
H & C & C & C & C & H & H & C & H
\\ \hline

Node 3 & H & H & C & C & C & C & C & H & H & \fbox{\textbf{F}} & H &
H & H & H & C & C & C & C & C & C
\\ \hline

Node 4 & C & C & C & C & \fbox{\textbf{M}} & C & H & H & C &
\fbox{\textbf{F}} & C & C & \fbox{\textbf{M}} & C & C & H &
\fbox{\textbf{M}} & C & C & \fbox{\textbf{M}}
\\ \hline

Node 5 & H & C & H & H & H & C & C & C & C & C & C & H & C & C & H &
C & C & H & C & H
\\ \hline

Node 6 & \fbox{\textbf{M}} & C & C & H & \fbox{\textbf{M}} & C & C &
\fbox{\textbf{F}} & H & \fbox{\textbf{F}} & C & C & H & H & C & H &
H & H & H & C
\\ \hline

Node 7 & H & H & C & C & H & C & H & C & C & \fbox{\textbf{F}} & H &
C & H & \fbox{\textbf{F}} & C & C & H & H & H & C
\\ \hline

Node 8 & C & H & H & H & H & H & H & H & C & C & H & H & H &
\fbox{\textbf{F}} & H & C & H & H & H & H
\\ \hline

\end{tabular}}
\vspace{0.1cm}
\end{table}
\setlength{\tabcolsep}{1.3pt}
\begin{table}[t]
\caption{Detection results for deployment scenario 2 in Fig.
\ref{fig:rst_numerical_example_sub_c}}\label{table:detection_2}\centering
\vspace{0.1cm}
\subtable[RST-based approach ($R_0 = 1 \ \mathrm{km}$)]{
   \label{table:detection_2_rst_1}
\begin{tabular}{c||cccccccccccccccccccc}\hline
Time slot & 1 & 2 & 3 & 4 & 5 & 6 & 7 & 8 & 9 & 10 & 11 & 12 & 13 &
14 & 15 & 16 & 17 & 18 & 19 & 20 \\ \hline\hline

Node 1 & C & C & C & C & C & C & H & C & H & H & C & H & H & C & H &
H & H & H & C & H
\\ \hline

Node 2 & H & C & H & H & C & \fbox{\textbf{F}} & C & H & C & C & H &
C & C & \fbox{\textbf{F}} & H & C & C & H & C & C
\\ \hline

Node 3 & H & C & H & H & H & H & H & \fbox{\textbf{M}} & C & C & C &
C & H & H & C & C & H & C & C & C
\\ \hline

Node 4 & H & C & H & H & H & H & C & C & H & H & C &
\fbox{\textbf{M}} & C & C & H & H & C & C & C & H
\\ \hline

Node 5 & H & C & C & H & H & H & \fbox{\textbf{F}} & H & H & C & H &
H & C & H & C & C & H & C & H & H
\\ \hline

Node 6 & H & C & C & H & C & H & H & H & H & H & H & H & H & C & C &
\fbox{\textbf{F}} & C & H & C & C
\\ \hline

Node 7 & C & H & C & H & H & C & C & C & H & H & H & C & C & C & H &
H & H & C & C & C
\\ \hline

Node 8 & H & C & H & C & C & H & C & H & C & C & H & H & H & C & H &
H & H & H & H & H
\\ \hline

\end{tabular}}
\vspace{0.3cm} \setlength{\tabcolsep}{1.03pt}
\subtable[RST-based approach ($R_0 = 0.5 \ \mathrm{km}$)]{
   \label{table:detection_2_rst_2}
\begin{tabular}{c||cccccccccccccccccccc}\hline
Time slot & 1 & 2 & 3 & 4 & 5 & 6 & 7 & 8 & 9 & 10 & 11 & 12 & 13 &
14 & 15 & 16 & 17 & 18 & 19 & 20 \\ \hline\hline

Node 1 & C & C & C & C & C & C & H & C & H & H & C & H & H & C & H &
H & H & H & C & H
\\ \hline

Node 2 & H & C & H & H & C & C & C & H & C & C & H & C & C & C & H &
C & C & H & C & C
\\ \hline

Node 3 & H & C & H & \fbox{\textbf{M}} & \fbox{\textbf{M}} & H & H &
\fbox{\textbf{M}} & C & C & C & C & H & H & C & C &
\fbox{\textbf{M}} & C & C & C
\\ \hline

Node 4 & H & C & H & \fbox{\textbf{M}} & \fbox{\textbf{M}} &
\fbox{\textbf{M}} & C & C & H & H & C & \fbox{\textbf{M}} & C & C &
H & \fbox{\textbf{M}} & C & C & C & H
\\ \hline

Node 5 & H & C & C & H & H & H & \fbox{\textbf{F}} & H & H & C & H &
H & C & H & C & C & H & C & H & H
\\ \hline

Node 6 & H & C & C & H & C & H & H & H & H & H & H & H & H & C & C &
C & C & H & C & C
\\ \hline

Node 7 & C & H & C & H & H & C & C & C & H & H & H & C & C & C & H &
H & H & C & C & C
\\ \hline

Node 8 & H & C & H & C & C & H & C & H & C & C & H & H & H & C & H &
H & H & H & H & H
\\ \hline

\end{tabular}}
\vspace{0.3cm} \setlength{\tabcolsep}{1.28pt}
\subtable[Classical approach using a bank of matched filters]{
   \label{table:detection_2_mf}
\begin{tabular}{c||cccccccccccccccccccc}\hline
Time slot & 1 & 2 & 3 & 4 & 5 & 6 & 7 & 8 & 9 & 10 & 11 & 12 & 13 &
14 & 15 & 16 & 17 & 18 & 19 & 20 \\ \hline\hline

Node 1 & C & C & C & C & C & C & H & C & H & H & C & H & H & C & H &
H & H & H & C & H
\\ \hline

Node 2 & H & C & H & H & C & \fbox{\textbf{F}} & C & H & C & C & H &
C & C & \fbox{\textbf{F}} & H & C & C & H & C & C
\\ \hline

Node 3 & H & C & H & H & H & H & H & \fbox{\textbf{M}} & C & C & C &
C & H & H & C & C & H & C & C & C
\\ \hline

Node 4 & H & C & H & H & H & H & C & C & H & H & C &
\fbox{\textbf{M}} & C & C & H & H & C & C & C & H
\\ \hline

Node 5 & H & C & C & H & H & H & \fbox{\textbf{F}} & H & H & C & H &
H & C & H & C & C & H & C & H & H
\\ \hline

Node 6 & H & \fbox{\textbf{F}} & C & H & C & H & H & H & H & H & H &
H & H & C & C & \fbox{\textbf{F}} & C & H & C & C
\\ \hline

Node 7 & C & H & C & H & H & C & C & C & H & H & H & C & C &
\fbox{\textbf{F}} & H & H & H & C & C & C
\\ \hline

Node 8 & H & \fbox{\textbf{F}} & H & C & C & H & C & H & C & C & H &
H & H & C & H & H & H & H & H & H
\\ \hline

\end{tabular}}
\vspace{0.1cm}
\end{table}

Table \ref{table:detection_1} and \ref{table:detection_2} show the
detection results for deployment scenario 1 and 2 in Fig.
\ref{fig:rst_numerical_example_sub_b}-(c), respectively. The letters
H, C, F, and M stand for \textit{hit}, \textit{correct rejection},
\textit{false alarm}, and \textit{miss}, respectively. Note that F
and M are the erroneous detections.
%
%
From Table \ref{table:detection_1}, it can be seen that a total of 1
\textit{miss} and 3 \textit{false alarms} are induced under the
RST-based approach with $R_0=1 \ \mathrm{km}$ (which is the entire
region of interest), and 2 \textit{misses} and 1 \textit{false
alarm} are occurred with $R_0=0.5 \ \mathrm{km}$. Under the
classical approach, however, a total of 6 \textit{misses} and 10
\textit{false alarms} occur, which is 4 times more than the
RST-based approach with $R_0 = 1 \ \mathrm{km}$ and 5.3 times more
than that with $R_0 = 0.5 \ \mathrm{km}$.

From Table \ref{table:detection_2}, we can see that a total of 2
\textit{misses} and 4 \textit{false alarms} are induced under the
RST-based approach with $R_0=1 \ \mathrm{km}$, and 9 \textit{misses}
and 1 \textit{false alarm} are occurred with $R_0=0.5 \
\mathrm{km}$. Overall, by reducing the discovery range $R_0$, the
occurrence of \textit{false alarm} is reduced, whereas that of
\textit{miss} is increased. Note, however, that most of the
\textit{misses} are due to the nodes outside the discovery region
(i.e., node 3 and 4). In fact, the detection of the nodes inside the
discovery region is more accurate than before. These are because the
density $f_{J}(\cdot)$ gives more weight to the smaller size of the
set $\mathbf{X}_t$ during the decision process. Also, it should be
mentioned that the nodes outside the discovery region can be
detected in a particular reception since there is no absolute and
deterministic boundary for a node detection. On the other hand,
under the classical approach, a total of 2 \textit{misses} and 7
\textit{false alarms} occur, which is 1.5 times more than the
RST-based approach with $R_0 = 1 \ \mathrm{km}$ and is tantamount to
that with $R_0 = 0.5 \ \mathrm{km}$. Note again that most of the
errors under the RST-based approach, specifically the
\textit{misses}, come from detecting the nodes outside the discovery
region which is actually a \textit{preferred} error.

\section{Concluding Remarks}\label{secCon}

In this paper, we studied the problem of neighbor discovery in a
wireless sensor network. By incorporating physical layer parameters,
we enabled a more accurate and realistic performance assessment of
the chosen neighbor discovery algorithm. Unlike the collision
channel, such incorporation required us to explicitly specify the
set of transmitting neighbors in each time slot based on the
received signal. With the aid of the theory of random set, we were
able to present an alternative method to the classical approach
using a bank of matched filters for detecting the set of
transmitting neighbors. The performance gain of using this new
method comes in an additional cost of complexity. Several steps are
still needed to complete our work. To fully validate the advantages
of the alternative method, the performance evaluation needs to be
supplemented with additional simulations. Also, it should be noted
that we focused on discovering unidirectional links as in most of
the previous work. However, for routing and other important
functions of a network, bidirectional links simplify the network
operation. Therefore, it is of interest to develop a self-organizing
protocol which establishes bidirectional links in a distributed
manner.
%
%

%
%


\appendices

\section{Fundamentals of the Theory of Random Set}\label{appendix:rst}

Random set theory (RST) and its associated finite-set statistics
(FISST) are extensively studied in the book \textit{Mathematics of
Data Fusion} \cite{goodman:mathematics}. This section briefly
introduces the essentials of RST, and refer to
\cite{goodman:mathematics} and other companion publications
\cite{mahler:engineering, mahler:random, mahler:statistical,
ali:random, vihola:random} for more details. In RST, the
\textit{belief mass} of a random finite set $\mathbf{X}$ plays a
similar role to that of the cumulative distribution function of a
random variable, and is defined as
\begin{equation}\label{eqn:appendix_rst_1}
\begin{array}{l}
\beta_{\mathbf{X}}(\mathbf{C}) \triangleq \mathrm{Pr}\{\mathbf{X}
\subseteq \mathbf{C}\}
\end{array}
\end{equation}
where $\mathbf{C}$ is a closed subset of the space $\mathbb{S}$. For
example \cite{mahler:engineering}, if $\mathbf{X}=\{\mathbf{x}\}$,
i.e., a singleton, where $\mathbf{x}$ is a random vector,
$\beta_{\mathbf{X}}(\mathbf{C}) = \mathrm{Pr}\{\mathbf{X} \subseteq
\mathbf{C}\} = \mathrm{Pr}\{\mathbf{x} \in \mathbf{C}\}$ where
$\mathrm{Pr}\{\mathbf{x} \in \mathbf{C}\}$ is the probability
measure on $\mathbb{S}$. From this, it can be conjectured that the
belief mass generalizes the ordinary probability measure. It is
straightforward to write the belief mass in
(\ref{eqn:appendix_rst_1}) as
\begin{equation}\label{eqn:appendix_rst_3}
\begin{array}{l}
\beta_{\mathbf{X}}(\mathbf{C}) = \displaystyle \sum_{\mathbf{B}
\subseteq \mathbf{C}} \mathrm{Pr}\{\mathbf{X}=\mathbf{B}\} =
\sum_{\mathbf{B} \subseteq \mathbf{C}} f_{\mathbf{X}}(\mathbf{B})
\end{array}
\end{equation}
where $f_{\mathbf{X}}(\cdot)$ is the \textit{belief density} of a
random set $\mathbf{X}$, and it plays the role of a probability
density function. One natural question is how to derive
$f_{\mathbf{X}}(\cdot)$ from $\beta_{\mathbf{X}}(\cdot)$ which will
be answered in the sequel.
%
%

Consider the case where the hybrid space $\mathbb{S}$ is comprised
only of a finite discrete space $U$. Then, the belief density
$f_{\mathbf{X}}(\cdot)$ of a random finite set $\mathbf{X}$ can be
obtained via the M\"{o}bius inverse transform of
$\beta_{\mathbf{X}}(\cdot)$ as
\begin{equation}\label{eqn:appendix_rst_4}
f_{\mathbf{X}}(\mathbf{B}) = \sum_{\mathbf{C}\subseteq \mathbf{B}}
(-1)^{|\mathbf{B} \setminus \mathbf{C}|}
\beta_{\mathbf{X}}(\mathbf{C})
\end{equation}
by viewing sets as points in another space.
%
%
%
\begin{example}\label{example_rst1}
Take $\mathbb{S}=U=\{a,b\}$. Then,
$\mathcal{P}(\mathbb{S})=\{\emptyset, \{a\}, \{b\}, \{a,b\}\}$.
Assign probability to each element of the power set so that the sum
is equal to 1:
\begin{displaymath}
f_{\mathbf{X}}(\emptyset) = 0.1, f_{\mathbf{X}}(\{a\}) = 0.4,
f_{\mathbf{X}}(\{b\}) = 0.3, f_{\mathbf{X}}(\{a,b\}) = 0.2
\end{displaymath}
Then, we can obtain the values of the belief mass using
(\ref{eqn:appendix_rst_3}). For example, $\beta_{\mathbf{X}}(\{b\})
= f_{\mathbf{X}}(\emptyset) + f_{\mathbf{X}}(\{b\}) = 0.4$.
Likewise, we have
\begin{displaymath}
\beta_{\mathbf{X}}(\emptyset) = 0.1, \beta_{\mathbf{X}}(\{a\}) =
0.5, \beta_{\mathbf{X}}(\{a,b\}) = 1
\end{displaymath}
We can also retrieve the values of the belief density from the
belief mass using (\ref{eqn:appendix_rst_4}). For example,
$f_{\mathbf{X}}(\{b\}) = - \beta_{\mathbf{X}}(\emptyset) +
\beta_{\mathbf{X}}(\{b\}) = 0.3$.
\end{example}
%
%
%
For a general case where $\mathbb{S}\triangleq \mathbb{R}^d \times
U$ with $d>0$, the M\"{o}bius inverse transform is not applicable
because it applies only for a finite partially ordered set
\cite{goodman:mathematics}. The continuous analog of the M\"{o}bius
inverse transform, which is often called the \textit{set
derivative}, at $\mathbf{Z}=\{\mathbf{z}_1, \dots, \mathbf{z}_n\}$
with $\mathbf{z}_1 \neq \dots \neq \mathbf{z}_n$, is defined by
\cite{mahler:random}
\begin{equation*}
    \displaystyle \frac{\delta F(\mathbf{C})}{\delta \mathbf{Z}}
    \triangleq \frac{\delta^n F}{\delta \mathbf{z}_n \cdots \delta
    \mathbf{z}_1}(\mathbf{C}) \triangleq \frac{\delta}{\delta
    \mathbf{z}_n} \frac{\delta^{n-1}F}{\delta \mathbf{z}_{n-1} \cdots
    \delta \mathbf{z}_1} (\mathbf{C})
\end{equation*}
where
\begin{equation*}
    \displaystyle \frac{\delta F(\mathbf{C})}{\delta \mathbf{z}_i}
    \triangleq \lim_{\nu(E_{\mathbf{z}_i})\rightarrow0}
    \frac{F(\mathbf{C}\cup E_{\mathbf{z}_i}) -
    F(\mathbf{C})}{\nu(E_{\mathbf{z}_i})}
\end{equation*}
where $E_{\mathbf{z}_i}$ is a small neighborhood of $\mathbf{z}_i$
and $\nu(\cdot)$ is the hyper-volume (i.e., Lebesgue measure) of a
given set, and $\frac{\delta F(\mathbf{C})}{\delta \emptyset}
\triangleq F(\mathbf{C})$.


\section{Derivation of (\ref{eqn:m-slot-correlated-1})}\label{appendix:derivation_of_correlated_case}

The details in deriving (\ref{eqn:m-slot-correlated-1}) is delivered
in this section. By applying Bayes rule to the conditional
probability, we have
\begin{multline}\label{eqn:appendix-B-1}
\displaystyle \mathrm{Pr} \Set{ k \in \mathbf{I}_t^s |\ \!\!
|\mathbf{I}_t^s|=h_t, r_k=r' }\\ = \frac{\mathrm{Pr} \{ k \in
\mathbf{I}_t^s, |\mathbf{I}_t^s|=h_t, r_k=r' \} }{ \mathrm{Pr} \{
|\mathbf{I}_t^s|=h_t, r_k=r' \}}
\end{multline}
whose numerator and denominator are specified one by one in the
following. Conditioning the numerator on the number of transmitters
and applying Bayes rule once again yields
\begin{multline}\label{eqn:appendix-B-2-temp}
\displaystyle \mathrm{Pr} \{ k \in \mathbf{I}_t^s,
|\mathbf{I}_t^s|=h_t, r_k=r' \} \\
= \displaystyle \sum_{n=h_t}^{J} \mathrm{Pr} \{ |\mathbf{I}_t^s|=h_t | k \in \mathbf{I}_t^s, r_k=r', |\mathbf{I}_t|=n \}\\
\cdot \mathrm{Pr} \{k \in \mathbf{I}_t^s, r_k=r', |\mathbf{I}_t|=n
\}
\end{multline}
Define $f_{1,n}$ as the probability that a transmitter located at
distance $r'$ from the receiver will succeed among $n(\geq1)$
simultaneous transmissions and it is given by
\begin{align*}\label{eqn:correlated-element-prob1}
f_{1,n} = 1 - \! \int_0^{\infty} \!\!\! \cdots \int_0^{\infty} \!\!
F_P \! \left(\tau \sum_{i=2}^n x_i
\right|\left.\vphantom{\sum_{i=2}^n x_i} r' \! \right) \! dF_P(x_2)
\cdots dF_P(x_n)
\end{align*}
%
%
Also, define $f_{2,n}$ as the probability that a transmitter at an
arbitrary distance will succeed among $n(>1)$ simultaneous
transmissions given that one of the other transmitters is known to
be located at distance $r'$ from the receiver which is obtained by
\begin{equation*}\label{eqn:correlated-element-prob2}
\displaystyle f_{2,n} = \displaystyle 1 - \! \int_0^{\infty} \!
\!\!\! \cdots \int_0^{\infty}\!\!\! F_P \! \left(\! \tau \!
\sum_{i=2}^n x_i \! \right) \! dF_P(x_2|r') dF_P(x_3) \cdots
dF_P(x_n)
\end{equation*}
and $f_{2,1}=0$. Using these probabilities, we can compute the
numerator as
\begin{multline}\label{eqn:appendix-B-2}
\displaystyle \mathrm{Pr} \{k \in \mathbf{I}_t^s,
|\mathbf{I}_t^s|=h_t, r_k=r' \}\\
= \displaystyle \sum_{n=h_t}^{J} { n-1 \choose h_t-1}
f_{2,n}^{h_t-1} (1-f_{2,n})^{n-h_t}\\
\cdot { J-1 \choose n-1} p_{\mathrm{T}}^{n} (1-p_{\mathrm{T}})^{J-n}
f_{1,n} f_r(r')
\end{multline}

Similarly, the denominator can be expanded as
\begin{multline}\label{eqn:appendix-B-3}
\displaystyle { \mathrm{Pr} \{ |\mathbf{I}_t^s|=h_t, r_k=r' \}} \\
=\displaystyle \sum_{n=h_t}^{J} \mathrm{Pr} \{ |\mathbf{I}_t^s|=h_t
| r_k=r', |\mathbf{I}_t|=n \} \\
\cdot \mathrm{Pr} \{ r_k=r', |\mathbf{I}_t|=n \}
\end{multline}
Denote by $\mathcal{J}_n$ the set of elements in the power set
$\mathcal{P}(\mathcal{J})$ whose cardinality is equal to $n$, and
$\mathcal{J}_n^k(\subseteq \mathcal{J}_n)$ the set of elements
containing a specific node index $k$. For example, if $\mathcal{J} =
\{1,2,3\}$, then $\mathcal{P}(\mathcal{J}) = \{ \emptyset, \{ 1\},
\{ 2\}, \{ 3\}, \{1,2 \}, \{1,3 \}, \{2,3 \}, \{1,2,3 \} \}$,
$\mathcal{J}_2 = \{ \{1,2 \}, \{1,3 \}, \{2,3 \} \}$, and
$\mathcal{J}_2^1 = \{ \{1,2 \}, \{1,3 \}\}$. Using these notations,
the first probability in the summation of \eqref{eqn:appendix-B-3}
can be split into
\begin{align}\label{eqn:appendix-B-4}
\displaystyle & \mathrm{Pr} \{|\mathbf{I}_t^s|=h_t | r_k=r',
|\mathbf{I}_t|=n \} \nonumber \\
& = \displaystyle \sum_{\mathcal{S} \in \mathcal{J}_n^k} \mathrm{Pr}
\{ k \in \mathbf{I}_t^s, |\mathbf{I}_t^s|=h_t,
\mathbf{I}_t=\mathcal{S} | r_k=r', |\mathbf{I}_t|=n \} \nonumber \\
& \ \ \ \ \ \ + \sum_{\mathcal{S} \in \mathcal{J}_n^k} \mathrm{Pr}
\{ k \notin \mathbf{I}_t^s, |\mathbf{I}_t^s|=h_t,
\mathbf{I}_t=\mathcal{S} |
r_k=r', |\mathbf{I}_t|=n \} \nonumber \\
& \ \ \ + \sum_{\mathcal{S} \in \mathcal{J}_n \setminus
\mathcal{J}_n^k} \mathrm{Pr} \{ |\mathbf{I}_t^s|=h_t,
\mathbf{I}_t=\mathcal{S} | r_k=r', |\mathbf{I}_t|=n \}
\end{align}
Since nodes are randomly transmitting with equal probability, each
realization of the set of transmitters having same cardinality are
equiprobable, i.e., $\mathrm{Pr}\Set{\mathbf{I}_t = \mathcal{S} |\
\!\! |\mathbf{I}_t|=n} = {1}/{{J \choose n}}$, for all $\mathcal{S}
\in \mathcal{J}_n$, and the set of transmitters $\mathbf{I}_t$
itself is independent over time slots (but the set of successful
transmitters $\mathbf{I}_t^s$ is not). Therefore, the probability in
the first summation of \eqref{eqn:appendix-B-4} is computed as
\begin{multline}
\displaystyle \mathrm{Pr} \{ k \in \mathbf{I}_t^s,
|\mathbf{I}_t^s|=h_t, \mathbf{I}_t=\mathcal{S} | r_k=r',
|\mathbf{I}_t|=n \} \\
= { n-1 \choose h_t-1} f_{2,n}^{h_t-1} (1-f_{2,n})^{n-h_t} f_{1,n}
\frac{1}{{J \choose n}}
\end{multline}
%
for all $\mathcal{S} \in \mathcal{J}_n^k$. Likewise, we can specify
the remaining probabilities in \eqref{eqn:appendix-B-4}. By noting
that $|\mathcal{J}_n|={J \choose n}$ and $|\mathcal{J}_n^k| = {J-1
\choose n-1}$ and after some manipulation, \eqref{eqn:appendix-B-4}
becomes
%
\begin{multline}\label{eqn:appendix-B-7}
\displaystyle \mathrm{Pr} \{ |\mathbf{I}_t^s|=h_t | r_k=r',
|\mathbf{I}_t|=n \}\\
= \displaystyle \frac{n}{J} { n-1 \choose h_t-1} f_{2,n}^{h_t-1}
(1-f_{2,n})^{n-h_t} f_{1,n}\\
+ \displaystyle \frac{n}{J} { n-1 \choose h_t} f_{2,n}^{h_t}
(1-f_{2,n})^{n-h_t-1} (1-f_{1,n})\\
+ \displaystyle \left(1-\frac{n}{J}\right) { n \choose h_t}
f_n^{h_t} (1-f_n)^{n-h_t}
\end{multline}
where $f_n$ is the shorthand notation for (\ref{eqn:capture-cal}).
The second probability in (\ref{eqn:appendix-B-3}) is simply
\begin{equation}\label{eqn:appendix-B-8}
\mathrm{Pr} \{ r_k=r', |\mathbf{I}_t|=n \} = {J \choose n}
{p_{\mathrm{T}}}^{n} (1-p_{\mathrm{T}})^{J-n}f_r(r')
\end{equation}
By substituting (\ref{eqn:appendix-B-7}) and
(\ref{eqn:appendix-B-8}) into (\ref{eqn:appendix-B-3}), the
denominator becomes
\begin{align}\label{eqn:appendix-B-9}
\displaystyle & \mathrm{Pr} \{ |\mathbf{I}_t^s|=h_t, r_k=r' \} \nonumber \\
& = \sum_{n=h_t}^{J} \left( \frac{n}{J} \left(\zeta_n + \gamma_n -
\xi_n\right) + \xi_n \right) {J \choose n} {p_{\mathrm{T}}}^n (1 -
p_{\mathrm{T}})^{J-n}f_r(r')
\end{align}
%
where the shorthand notations $\zeta_n$, $\gamma_n$ and $\xi_n$ are
defined in Section \ref{sec:multslot}.

\bibliographystyle{IEEEtran}
\bibliography{bib_jh_nd}

\begin{thebibliography}{10}
\providecommand{\url}[1]{#1}
\csname url@samestyle\endcsname
\providecommand{\newblock}{\relax}
\providecommand{\bibinfo}[2]{#2}
\providecommand{\BIBentrySTDinterwordspacing}{\spaceskip=0pt\relax}
\providecommand{\BIBentryALTinterwordstretchfactor}{4}
\providecommand{\BIBentryALTinterwordspacing}{\spaceskip=\fontdimen2\font plus
\BIBentryALTinterwordstretchfactor\fontdimen3\font minus
  \fontdimen4\font\relax}
\providecommand{\BIBforeignlanguage}[2]{{%
\expandafter\ifx\csname l@#1\endcsname\relax
\typeout{** WARNING: IEEEtran.bst: No hyphenation pattern has been}%
\typeout{** loaded for the language `#1'. Using the pattern for}%
\typeout{** the default language instead.}%
\else
\language=\csname l@#1\endcsname
\fi
#2}}
\providecommand{\BIBdecl}{\relax}
\BIBdecl

\bibitem{borbash:asynchronous}
S.~A. Borbash, A.~Ephremides, and M.~J. McGlynn, ``An asynchronous neighbor
  discovery algorithm for wireless sensor networks,'' \emph{Ad Hoc Networks},
  vol.~5, pp. 998--1016, 2007.

\bibitem{vasudevan:coupon}
S.~Vasudevan, D.~Towsley, D.~Goeckel, and R.~Khalili, ``Neighbor discovery in
  wireless networks and the coupon collector's problem,'' in \emph{ACM
  MobiCom}, Beijing, China, Sep. 2009.

\bibitem{vasudevan:neighbor}
S.~Vasudevan, J.~Kurose, and D.~Towsley, ``On neighbor discovery in wireless
  networks with directional antennas,'' in \emph{IEEE INFOCOM}, Miami, Florida,
  Mar. 2005.

\bibitem{mcglynn:birthday}
M.~J. McGlynn and S.~A. Borbash, ``Birthday protocols for low energy deployment
  and flexible neighbor discovery in ad hoc wireless networks,'' in \emph{ACM
  MobiHoc}, Long Beach, California, Oct. 2001.

\bibitem{zorzi:capture}
M.~Zorzi and R.~Rao, ``Capture and retransmission control in mobile radio,''
  \emph{IEEE Journal on Selected Areas in Communications}, vol.~12, no.~8, pp.
  1289 -- 1298, Oct. 1994.

\bibitem{nguyen:capture}
G.~D. Nguyen, A.~Ephremides, and J.~E. Wieselthier, ``On capture in
  random-access systems,'' in \emph{IEEE ISIT}, Seattle, Washington, Jul. 2006.

\bibitem{namislo:analysis}
C.~Namislo, ``Analysis of mobile radio slotted aloha networks,'' \emph{IEEE
  Journal on Selected Areas in Communications}, vol. SAC-2, no.~4, pp. 583 --
  588, Jul. 1984.

\bibitem{hajek:capture}
B.~Hajek, A.~Krishna, and R.~O. LaMaire, ``On the capture probability for a
  large number of stations,'' \emph{IEEE Transactions on Communications},
  vol.~45, no.~2, pp. 254--260, 1997.

\bibitem{ghez:stability}
S.~Ghez and S.~Verd\'{u}, ``Stability property of slotted aloha with
  multipacket reception capability,'' \emph{IEEE Transactions on Automatic
  Control}, vol.~33, no.~7, pp. 640 -- 649, Jul. 1988.

\bibitem{goodman:mathematics}
I.~R. Goodman, R.~P.~S. Mahler, and H.~T. Nguyen, \emph{Mathematics of Data
  Fusion}.\hskip 1em plus 0.5em minus 0.4em\relax Dordrecht, The Netherlands:
  Kluwer Academic Publishers, 1997.

\bibitem{ali:random}
A.~M. Ali, R.~Hudson, F.~Lorenzelli, and K.~Yao, \emph{A Random Finite Set
  Approach to Joint Estimation/Detection/Tracking/Fusion in a Wireless Sensor
  Network}.\hskip 1em plus 0.5em minus 0.4em\relax University of California,
  Los Angeles: Research Report, 2008.

\bibitem{vihola:random}
M.~Vihola, \emph{Random Sets for Multitarget Tracking and Data Fusion}.\hskip
  1em plus 0.5em minus 0.4em\relax Tampere University of Technology: Licentiate
  Thesis, 2004.

\bibitem{mahler:engineering}
R.~P.~S. Mahler, ``Engineering statistics for multi-object tracking,'' in
  \emph{IEEE Workshop on Multi-Object Tracking}, Vancouver, Canada, Jul. 2001.

\bibitem{mahler:random}
------, ``Random sets: Unification and computation for information fusion--a
  retrospective assessment,'' in \emph{The 7th International Conference on
  Information Fusion}, Stockholm, Sweden, Jul. 2004.

\bibitem{mahler:statistical}
------, \emph{Statistical Multisource-Multitarget Information Fusion}.\hskip
  1em plus 0.5em minus 0.4em\relax Norwood, MA: Artech House, 2007.

\bibitem{biglieri:multiuser}
E.~Biglieri and M.~Lops, ``Multiuser detection in a dynamic environment-part 1:
  User identification and data detection,'' \emph{IEEE Transactions on
  Information Theory}, vol.~53, no.~9, pp. 3158 -- 3170, Sep. 2007.

\bibitem{tse:fundamentals}
D.~Tse and P.~Viswanath, \emph{Fundamentals of Wireless Communication}.\hskip
  1em plus 0.5em minus 0.4em\relax Cambridge, United Kingdom: Cambridge
  university press, 2005.

\bibitem{goldsmith:wireless}
A.~Goldsmith, \emph{Wireless Communications}.\hskip 1em plus 0.5em minus
  0.4em\relax Cambridge, United Kingdom: Cambridge university press, 2005.

\bibitem{leon:probability}
A.~Leon-Garcia, \emph{Probability and Random Processes for Electrical
  Engineering}, 2nd~ed.\hskip 1em plus 0.5em minus 0.4em\relax Reading, MA:
  Addison-Wesley Publishing Company, 1994.

\bibitem{poor:introduction}
H.~V. Poor, \emph{An Introduction to Signal Detection and Estimation},
  2nd~ed.\hskip 1em plus 0.5em minus 0.4em\relax New York: Springer-Verlag,
  1994.

\bibitem{xbow:micaz}
\BIBentryALTinterwordspacing
{XBOW MICAz Mote Specifications}. [Online]. Available:
  \url{http://www.xbow.com}
\BIBentrySTDinterwordspacing

\bibitem{stallings:wireless}
W.~Stallings, \emph{Wireless Communications and Networks}, 2nd~ed.\hskip 1em
  plus 0.5em minus 0.4em\relax Upper Saddle River, NJ: Pearson Prentice Hall,
  2005.

\end{thebibliography}

\end{document}